\documentclass[showpacs,amsmath,amssymb,aps,preprint,nofootinbib]{revtex4-1}

\usepackage{graphicx}
\usepackage{dcolumn}
\usepackage{color}
\usepackage{float}
\usepackage{hyperref}
\hypersetup{%
colorlinks=true,%
linkcolor=blue,
citecolor=blue,
urlcolor=magenta
}
\newcommand{\sumint}{\mbox{$\sum$}\kern-2.7ex\int}

\newcommand{\ztwosla}{\text{Z}\kern-0.8ex /}

\def\gtsim{\mathrel{\hbox{\raise0.2ex
\hbox{$>$}\kern-0.75em\raise-0.9ex\hbox{$\sim$}}}}
\def\ltsim{\mathrel{\hbox{\raise0.2ex
\hbox{$<$}\kern-0.75em\raise-0.9ex\hbox{$\sim$}}}}

\begin{document}


\title{Gravitational waves from CP domain wall collapse and electron EDM in a complex singlet model with dimension-five Yukawa interactions}

\author{Hieu The Pham$^{1}$}
\email{hieupham@gapp.nthu.edu.tw}
\author{Eibun Senaha$^{2,3}$}
\email{eibunsenaha@vlu.edu.vn}
\affiliation{$^1$ Department of Physics, National Tsing Hua University, Hsinchu 300, Taiwan}
\affiliation{$^2$Subatomic Physics Research Group, Science and Technology Advanced Institute, Van Lang University, Ho Chi Minh City, Vietnam}
\affiliation{$^3$Faculty of Applied Technology, Van Lang School of Technology, Van Lang University, Ho Chi Minh City, Vietnam}

\date{\today}

\begin{abstract}
We study the interplay between gravitational waves (GWs) from domain wall collapse and the electron electric dipole moment (EDM) in a complex singlet extension of the standard model with dimension-five Yukawa interactions. 
In this framework, the scalar potential admits CP-related degenerate vacua, leading to the formation of CP domain walls. 
While the resulting GW signal provides a probe of the vacuum structure of the singlet scalar sector, it does not by itself constitute a CP-violating observable. 
Once the singlet scalar is coupled to standard model fermions, CP-violating phases become observable through EDMs. 
We analyze whether current and future EDM experiments can probe the parameter region where the GW signal is detectable by SKA and THEIA. 
We find that the current electron EDM bound already constrains part of the parameter space, while future sensitivities at the level of $10^{-31}$--$10^{-32}\,e\,\mathrm{cm}$ can probe regions overlapping with the GW-detectable domain. 
Our results highlight the complementarity between GW and EDM observables in probing the singlet scalar sector, providing a coherent picture of its vacuum structure and CP properties.
\end{abstract}


\maketitle


\section{Introduction}\label{sec:intro}

Although the Standard Model (SM) is in excellent agreement with data from terrestrial experiments, it fails to explain several cosmological observations, such as the baryon asymmetry of the Universe and dark matter, suggesting the existence of physics beyond the SM.
Among the many possible extensions, enlarging the Higgs sector is one of the most widely studied possibilities, and new sources of CP violation (CPV) are likely to exist unless some mechanism protects CP symmetry.

A complex singlet scalar extension of the SM (cxSM)~\cite{Barger:2008jx} is one of the simplest frameworks accommodating rich phenomenology, including Higgs physics~\cite{Costa:2014qga,Azevedo:2018llq,Abe:2021nih,Egle:2022wmq,Egle:2023pbm,Cho:2022zfg,Zhang:2023mnu}, dark matter~\cite{Barger:2010yn,Gonderinger:2012rd}, a strong first-order electroweak phase transition~\cite{Chiang:2017nmu,Cho:2021itv,Cho:2022our,Liang:2025zku,Biekotter:2025npc,Ghosh:2025rbt} and its associated gravitational wave (GW) signals~\cite{Cho:2022our,Liang:2025zku,Biekotter:2025npc,Ghosh:2025rbt}, as well as GWs from domain wall (DW) collapse~\cite{Chen:2020wvu,Pham:2024vso}.

As studied in Ref.~\cite{Chen:2020wvu}, the cxSM can accommodate the existence of CP domain walls (CPDWs) associated with multiple CP-related vacua $v_S$ and $v_S^*$. 
Here $v_S$ denotes the vacuum expectation value (VEV) of the singlet scalar field $S$, and the scalar potential is invariant under the transformation $S \to S^*$, or equivalently $\mathrm{Im}\,S \to -\,\mathrm{Im}\,S$. 
GWs generated by the collapse of such DWs have been proposed as a probe of CPV in the singlet scalar sector.
Although this probe provides a unique window into the vacuum structure of the singlet scalar sector, $\mathrm{Im}\,v_S$ is not directly related to the conventional CPV probed in laboratory experiments, since the singlet scalar field does not couple directly to SM fermions and hence does not induce CPV observables such as electric dipole moments (EDMs). 
In particular, the GW signal itself is not a CPV observable. 
Therefore, interpreting the GW signal as a manifestation of spontaneous CPV (SCPV) in the singlet scalar sector remains indirect. 
To render the singlet scalar phase physically observable as a CPV phase, the model must therefore be extended.\footnote{For realization of an electroweak baryogenesis~\cite{Kuzmin:1985mm,Rubakov:1996vz,*Funakubo:1996dw,*Riotto:1998bt,*Trodden:1998ym,*Bernreuther:2002uj,*Cline:2006ts,*Morrissey:2012db,*Konstandin:2013caa,*Senaha:2020mop} in the cxSM with higher-dimensional operators, see, e.g., Refs~\cite{Jiang:2015cwa,Grzadkowski:2018nbc,Idegawa:2023bkh}.}

In recent years, the upper bound on the electron EDM ($d_e$) has been significantly improved.
The current most stringent limit is $|d_e| < 4.1\times 10^{-30}~e\,\mathrm{cm}$ at the 90\% confidence level, set by the JILA Collaboration~\cite{Roussy:2022cmp}. 
This remarkable sensitivity suggests that the electron EDM can provide a powerful complementary probe of CPV in the singlet scalar sector once the model is suitably extended. 
A central question we address in this work is whether current or future electron EDM experiments can probe the parameter region where the GW signal is detectable. 
The combined observation of GWs and EDM effects would provide complementary evidence for CPV in the singlet scalar sector and lend support to the interpretation that the GW signal may originate from CPDW collapses.

In this paper, we extend the cxSM by introducing dimension-five Yukawa interactions that couple the singlet field to SM fermions and address the above question. 
In this framework, SCPV responsible for DW formation does not by itself generate observable CP violation, but can give rise to EDMs once the singlet scalar is coupled to fermions.
To elucidate the CPV structure of the extended model, we construct basis-invariant CP-odd quantities that characterize both explicit CPV and SCPV. 
We then derive the CPDW configurations associated with SCPV and study the interplay between GW signals from DW annihilation and the electron EDM. In particular, we investigate whether the parameter region accessible to pulsar timing arrays, such as SKA~\cite{Janssen:2014dka} and space-based detectors like THEIA~\cite{Theia:2017xtk}, can be probed by current or future EDM experiments.

Our analysis shows that the current electron EDM bound already excludes the parameter region in which the imaginary part of the singlet scalar VEV is below approximately $\mathcal{O}(10)$~TeV, near the maximal Higgs mixing allowed by current experiments. 
In contrast, the parameter region relevant for GW detection by SKA and THEIA corresponds to larger values of the singlet VEV, typically above $\mathcal{O}(10)$~TeV. 
We find that this region can be probed by future EDM experiments if the sensitivity reaches the $10^{-32}$--$10^{-31}$~$e\,\mathrm{cm}$ level, under mild assumptions on the dimension-five Yukawa couplings. 
This complementarity between cosmological and precision probes provides a novel strategy to explore the singlet scalar sector, linking its vacuum structure to observable CP-violating effects.

The paper is organized as follows. 
In Sec.~\ref{sec:model}, we define the Lagrangian of the cxSM with dimension-five Yukawa interactions, introduce CP-odd basis invariants, and derive the Higgs masses and mixings at tree level. 
In Sec.~\ref{sec:cpdw}, we analytically study SCPV and determine the vacuum phase. 
We then derive the equations of motion for the CPDW and present their solutions in a simplified setup, and discuss the bias terms responsible for DW collapse.
In Sec.~\ref{sec:gw}, we present analytic expressions for the GW spectrum and peak frequency from DW annihilation, while Sec.~\ref{sec:edm} discusses the connection to the electron EDM. In Sec.~\ref{sec:results}, we perform a full numerical analysis. 
Finally, Sec.~\ref{sec:conclusion} is devoted to conclusions and discussions.

\section{Model}\label{sec:model}
The cxSM extends the SM by introducing a complex scalar SU(2) singlet, $S$~\cite{Barger:2008jx}. 
In general, $S$ can acquire a VEV, leading to a rich vacuum structure. 
Correspondingly, diverse patterns of phase transitions can arise at finite temperature. 
In particular, a strong first-order phase transition~\cite{Chiang:2017nmu,Cho:2021itv,Cho:2022our,Liang:2025zku,Biekotter:2025npc,Ghosh:2025rbt} is phenomenologically attractive, as it provides one of the necessary ingredients for successful electroweak baryogenesis and can generate GW signals. 
Moreover, the spontaneous breaking of discrete symmetries can induce cosmological DWs, whose subsequent decay also produces GW signals~\cite{Chen:2020wvu,Pham:2024vso}. 
In addition to the phase transition physics, the model can accommodate a viable cold dark matter candidate~\cite{Barger:2010yn,Gonderinger:2012rd} if the scalar sector preserves CP.

The Lagrangian of the cxSM with a dimension-five Yukawa interaction is defined as
\begin{align}
\mathcal{L} = \mathcal{L}_\mathrm{cxSM}+\mathcal{L}_{h_i\bar{f}f}^\text{dim-5}.
\end{align}
As proposed in Ref.~\cite{Barger:2008jx}, we consider a minimal scalar potential
\begin{align}
V_0(H, S) = \frac{m^2}{2}H^\dagger H+\frac{\lambda}{4}(H^\dagger H)^2
	+\frac{\delta_2}{2}H^\dagger H|S|^2+\frac{b_2}{2}|S|^2+\frac{d_2}{4}|S|^4
	+\bigg(a_1S+\frac{b_1}{4}S^2+{\rm H.c.}\bigg),\label{V0}
\end{align}
where $a_1$ and $b_1$ are generally complex.
Both $a_1$ and $b_1$ break a global U(1), and furthermore, $a_1$ also breaks the $\mathbb{Z}_2$ symmetry $S\to -S$.
In the case that $a_1$ and $b_1$ are real, $V_0$ possesses a CP symmetry $S\to S^*$ or
a $\mathbb{Z}_2$ symmetry $\mathrm{Im}\,S\to -\mathrm{Im}\,S$.
This type of $\mathbb{Z}_2$ symmetry guarantees the stability of the dark matter. 

The Higgs doublet and singlet fields are parametrized as
\begin{align}
H(x) &=
	\left(
		\begin{array}{c}
		G^+(x) \\
		\frac{1}{\sqrt{2}}\big(v+h(x)+iG^0(x)\big)
		\end{array}
	\right),\\
S(x) &= \frac{1}{\sqrt{2}}\left(v_S + s(x)+i\chi(x) \right),
\end{align}
where $v$ and $v_S$ are the VEVs, and the former is given by
$v=1/\sqrt{\sqrt{2}G_F}\simeq 246.22$ GeV~\cite{ParticleDataGroup:2024cfk} with $G_F$ being a Fermi constant, while the latter can be generally complex, which is denoted as $v_S= v_S^r+iv_S^i=|v_S|e^{i\theta_{v_S}}$. $G^{0}(x)$ and $G^{+}(x)$ are the Nambu-Goldstone fields. 
When $a_1$ and $b_1$ are both real, and $v_S^i\neq0$, the aforementioned CP symmetry is spontaneously broken, generating a DW (referred to as CPDW~\cite{Chen:2020wvu}).

The Lagrangian of the Yukawa couplings, including the dimension-five Yukawa interaction, is defined as
\begin{align}
-\mathcal{L}_Y &= \bar{q}_{L}\tilde{H}\left(Y^u+\frac{C^u}{\Lambda}S\right)u_{R}
+ \bar{q}_{L}H\left(Y^d+\frac{C^d}{\Lambda}S\right)d_{R} +\bar{\ell}_{L}H\left(Y^e+\frac{C^e}{\Lambda}S\right)e_{R}+{\rm H.c.},
\end{align}
where $q_L = (u_L, d_L)^T$, $\ell_L = (\nu_L, e_L)^T$. $Y^f$ and $C^f$ are arbitrary three-by-three matrices and $\Lambda$ is a cutoff.
This effective theory may arise from integrating out heavy vector-like fermions in UV completions  (see, e.g, Ref.~\cite{Idegawa:2023bkh}).
Owing to these effective Yukawa interactions, $s$ and $\chi$ are defined as CP-even and CP-odd, respectively.
For each fermion sector $f=u,d,e$, $Y^f$ can be diagonalized by biunitary transformations of fermion fields $f_L =  V_L^ff_L'$ and $f_R= V_R^ff_R'$:
\begin{align}
V_L^{f\dagger}Y_\mathrm{SM}^f V_R^f& =Y^f_\mathrm{diag}=\mathrm{diag}(y_{f_1}, y_{f_2}, y_{f_3}), \\
V_L^{f\dagger}C^f V_R^f &= c^f,
\end{align}
where $Y^f_\mathrm{SM}  = Y^f +C^fv_S/(\sqrt{2}\Lambda)$, $c^f$ are general three-by-three complex matrices.
As shown in Appendix~\ref{basisCPV}, one of the CP-odd basis invariants, including $c^f$ can be defined as
\begin{align}
I^f &= \mathrm{Im}\mathrm{Tr}(Y_\mathrm{SM}^f a_1 C^{f\dagger}) = \mathrm{Im}\mathrm{Tr}(Y_\mathrm{diag}^f a_1 c^{f\dagger}) = \mathrm{Im}\left[a_1\sum_{i=1}^3(y_i^fc_{ii}^{f*})\right].
\end{align}
In this invariant, the off diagonal elements of $c^f$ do not appear.
For our purpose, we only focus on the top quark and electron contributions, $c_{33}^u\equiv c_t$, $c_{33}^e\equiv c_e$, and others $c^f$ are negligibly small.
For later use, we define the CP-odd invariants as
\begin{align}
I_1 & = \mathrm{Im}(a_1^2b_1^*) = |a_1|^2|b_1|\sin\theta_1, \\
I_2& = \mathrm{Im}(a_1y_tc_t^*) =|y_t||a_1||c_t|\sin\theta_2 , \\
I_3 &= \mathrm{Im}(a_1y_ec_e^*) =|y_e||a_1||c_e|\sin\theta_3, 
\end{align}
where $a_1 = |a_1|e^{i\theta_{a_1}}$, $b_1 = |b_1|e^{i\theta_{b_1}}$, $\theta_1 = 2\theta_{a_1}-\theta_{b_1}$, and $\theta_2 = \theta_{y_t}+\theta_{a_1}-\theta_{c_t}$, and $\theta_3 = \theta_{y_e}+\theta_{a_1}-\theta_{c_e}$. 
If $I_1 = I_2 = I_3 = 0$, there is no explicit CP violation. In this case, there exists a basis in which all parameters are real, which we refer to as a real basis.

Now, we consider the rephasing invariants for the SCPV.
 As detailed in Appendix~\ref{basisCPV}, the rephasing invariants, including $v_S$ are defined as
\begin{align}
J_a & = \mathrm{Im}(a_1v_S) = |a_1||v_S|\sin\phi_1, \\
J_b & = \mathrm{Im}(b_1v_S^{2}) = |b_1||v_S|^2\sin\phi_2,
\end{align}
where $\phi_1= \theta_{a_1}+\theta_{v_S}$ and $\phi_2= \theta_{b_1}+2\theta_{v_S}$.
Since $\phi_2 = 2\phi_1-\theta_1$, it is sufficient to consider $J_a$ for discussing SCPV.
The necessary and sufficient condition for SCPV is
\begin{align}
I_1 = I_2 = I_3=0\quad \mbox{and}\quad J_a \neq 0,
\end{align}
yielding $\sin\theta_{v_S}\neq0$ mod $\pi$.

The tadpole conditions with respect to $h$, $s$ and $\chi$ are, respectively, given by
\begin{align}
\left\langle \frac{\partial V_0}{\partial h}\right\rangle & = 
v\left[
\frac{m^2}{2} + \frac{\lambda}{4}v^2+\frac{\delta_2}{4}|v_S|^2 
\right]= 0,\label{tad_h}\\
\left\langle \frac{\partial V_0}{\partial s}\right\rangle & = 
v_S^r
\left[
\frac{b_2}{2}+\frac{\delta_2}{4}v^2+\frac{d_2}{4}|v_S|^2+\frac{b_1^r}{2}
\right]+\sqrt{2}a_1^r-\frac{b_1^i}{2}v_S^i
 = 0,\label{tad_s} \\
\left\langle \frac{\partial V_0}{\partial \chi}\right\rangle & = 
v_S^i
\left[
\frac{b_2}{2}+\frac{\delta_2}{4}v^2+\frac{d_2}{4}|v_S|^2-\frac{b_1^r}{2}
\right]-\sqrt{2}a_1^i-\frac{b_1^i}{2}v_S^r
= 0,\label{tad_chi}
\end{align}
where the symbol $\langle\cdots \rangle$ denotes that fluctuation fields are taken to be zero.
$a_1=a_1^r+ia_1^i=|a_1|\cos\theta_{a_1}+i|a_1|\sin\theta_{a_1}$ and $b_1=b_1^r+ib_1^i=|b_1|\cos\theta_{b_1}+i|b_1|\sin\theta_{b_1}$.
In our study, $v\neq0$, $v_S^r \neq 0$, and $v_S^i \neq 0$.
Note that (\ref{tad_s})$\times v_S^i-(\ref{tad_chi}) \times v_S^r$ is expressed in terms of $J_a$ and $J_b$:
\begin{align}
\sqrt{2}J_a + \frac{1}{2}J_b = 0.
\label{scpv_J}
\end{align}
In the real basis, Eq.~(\ref{scpv_J}) determines the magnitude of SCPV if it exists.
We will scrutinize it in Sec.~\ref{sec:cpdw}.

The mass matrix is defined as
\begin{align}
\frac{1}{2}
\begin{pmatrix}
h & s & \chi 
\end{pmatrix}
\mathcal{M}_S^2
\begin{pmatrix}
h \\ 
s \\
\chi 
\end{pmatrix}.
\end{align}
At the tree level, each mass matrix element is, respectively, given by
\begin{align}
(\mathcal{M}_S^2)_{11}&=\left\langle \frac{\partial^2 V_0}{\partial h^2}\right\rangle
= \frac{m^2}{2}+\frac{3\lambda}{4}v^2+\frac{\delta_2}{4}|v_S|^2,\\
(\mathcal{M}_S^2)_{22}&=\left\langle \frac{\partial^2 V_0}{\partial s^2}\right\rangle
= \frac{b_2}{2}+\frac{b_1^r}{2}+\frac{d_2}{4}(3v_S^{r2}+v_S^{i2})+\frac{\delta_2}{4}v^2,\\
(\mathcal{M}_S^2)_{33}&=\left\langle \frac{\partial^2 V_0}{\partial \chi^2}\right\rangle
= \frac{b_2}{2}-\frac{b_1^r}{2}+\frac{d_2}{4}(v_S^{r2}+3v_S^{i2})+\frac{\delta_2}{4}v^2,\\
(\mathcal{M}_S^2)_{12}&=\left\langle \frac{\partial^2 V_0}{\partial h\partial s}\right\rangle
= \frac{\delta_2}{2}vv_S^r,\\
(\mathcal{M}_S^2)_{13}&=\left\langle \frac{\partial^2 V_0}{\partial h\partial \chi}\right\rangle
= \frac{\delta_2}{2}vv_S^i,\\
(\mathcal{M}_S^2)_{23}&=\left\langle \frac{\partial^2 V_0}{\partial s\partial \chi}\right\rangle
= -\frac{b_1^i}{2}+\frac{d_2}{2}v_S^rv_S^i.
\end{align}
Note that $(\mathcal{M}_S^2)_{13}$ and $(\mathcal{M}_S^2)_{23}$ are nonzero if CPV exists.
Furthermore, we have the relation $(\mathcal{M}_S^2)_{13} = (\mathcal{M}_S^2)_{12} v_S^i/v_S^r$, reducing the degrees of freedom of $\mathcal{M}_S^2$ to 5.

Using the tadpole conditions (\ref{tad_h}), (\ref{tad_s}), and (\ref{tad_chi}) to eliminate $m^2$, $b_2$, and $b_1^r$, the mass matrix is reduced to
\begin{align}
\mathcal{M}_S^2 = 
\begin{pmatrix}
\frac{\lambda}{2}v^2 & \frac{\delta_2}{2}vv_S^r & \frac{\delta_2}{2}vv_S^i \\
\frac{\delta_2}{2}vv_S^r  & \frac{d_2}{2}v_S^{r2}-\frac{\sqrt{2}a_1^r}{v_S^r}+\frac{b_1^i}{2}\frac{v_S^i}{v_S^r} & -\frac{b_1^i}{2}+\frac{d_2}{2}v_S^rv_S^i \\
 \frac{\delta_2}{2}vv_S^i & -\frac{b_1^i}{2}+\frac{d_2}{2}v_S^rv_S^i & \frac{d_2}{2}v_S^{i^2}+\frac{\sqrt{2}a_1^i}{v_S^i}+\frac{b_1^i}{2}\frac{v_S^r}{v_S^i}.
\end{pmatrix}.\label{M}
\end{align}
We diagonalize $\mathcal{M}_S^2$ by an orthogonal matrix $O$ such that $O^T\mathcal{M}_S^2O = \mathrm{diag}(m_{h_1}^2, m_{h_2}^2, m_{h_3}^2)$,
where $O$ is parametrized as
\begin{align}
O & = 
\begin{pmatrix}
1 & 0 & 0 \\
0 & c_3 & -s_3 \\
0 & s_3 & c_3
\end{pmatrix}
\begin{pmatrix}
c_2 & 0 & -s_2  \\
0 & 1 & 0 \\
s_2 & 0 & c_2
\end{pmatrix}
\begin{pmatrix}
c_1 & -s_1 & 0 \\
s_1 & c_1 & 0 \\
0 & 0 & 1
\end{pmatrix},
\label{Omat}
\end{align}
with $s_i=\sin\alpha_i$ and $c_i=\cos\alpha_i~(i=1,2,3)$.

From $(\mathcal{M}_S^2)_{ij}=\sum_kO_{ik}O_{jk}m_{h_k}^2$, the quartic couplings in the scalar potential can be expressed as
\begin{align}
\lambda &= \frac{2}{v^2}\sum_iO_{1i}^2m_{h_i}^2, \label{lam} \\
\delta_2 & = \frac{2}{vv_S^r}\sum_iO_{1i}O_{2i}m_{h_i}^2=\frac{2}{vv_S^i}\sum_iO_{1i}O_{3i}m_{h_i}^2,\label{del2} \\
d_2 & = \frac{2}{v_S^{r2}}\left[\frac{\sqrt{2}a_1^r}{v_S^r}-\frac{b_1^i}{2}\frac{v_S^i}{v_S^r}+\sum_iO_{2i}^2m_{h_i}^2 \right]
= \frac{2}{v_S^{i2}}\left[-\frac{\sqrt{2}a_1^i}{v_S^i}-\frac{b_1^i}{2}\frac{v_S^r}{v_S^i}+\sum_iO_{3i}^2m_{h_i}^2 \right] \nonumber\\
& = \frac{2}{v_S^rv_S^i}\left[\frac{b_1^i}{2}+\sum_iO_{2i}O_{3i}m_{h_i}^2 \right]. \label{d2}
\end{align}
From the expressions of $d_2$, one can solve for $a_1^r$ and $a_1^i$ as
\begin{align}
a_1^r & = \frac{v_S^r}{\sqrt{2}}
\left[
\frac{b_1^i}{2}\frac{|v_S|^2}{v_S^rv_S^i}-\sum_iO_{2i}\left(O_{2i}-O_{3i}\frac{v_S^r}{v_S^i}\right)m_{h_i}^2
\right], \label{a1r}\\
a_1^i & = -\frac{v_S^i}{\sqrt{2}}
\left[
\frac{b_1^i}{2}\frac{|v_S|^2}{v_S^rv_S^i}-\sum_iO_{3i}\left(O_{3i}-O_{2i}\frac{v_S^i}{v_S^r}\right)m_{h_i}^2
\right].\label{a1i}
\end{align}

We now count the number of independent parameters in the scalar sector. 
The scalar potential contains five real parameters and two complex parameters. 
As discussed above, the phases of $a_1$ and $b_1$ are not independently physical; the rephasing-invariant phase is $\theta_1=2\theta_{a_1}-\theta_{b_1}$. In the absence of explicit CPV, we can work in a real basis with $a_1^i=b_1^i=0$, in which CPDWs can arise from SCPV. The scalar potential then contains seven real parameters:
\begin{align}
m^2,\quad \lambda,\quad \delta_2,\quad b_2,\quad d_2,\quad a_1^r,\quad b_1^r.
\end{align}
These parameters can be reconstructed from seven independent quantities chosen from the set
\begin{align}
v,\quad v_S^r,\quad v_S^i,\quad m_{h_1},\quad m_{h_2},\quad m_{h_3},\quad \alpha_1, \quad \alpha_2,\quad \alpha_3,
\end{align}
using the tadpole conditions, Eqs.~\eqref{tad_h}, \eqref{tad_s}, and \eqref{tad_chi}, together with Eqs.~\eqref{lam}, \eqref{del2}, \eqref{d2}, and \eqref{a1r}. The nine quantities are subject to two constraints:
\begin{align}
(\mathcal{M}_S^2)_{13}=(\mathcal{M}_S^2)_{12}\frac{v_S^i}{v_S^r},\quad
(\mathcal{M}_S^2)_{33} =(\mathcal{M}_S^2)_{32}\frac{v_S^i}{v_S^r},
\end{align}
where the second relation follows from Eq.~\eqref{a1i} with $a_1^i=b_1^i=0$.
For the CPDW analysis in Sec.~\ref{sec:cpdw}, we choose $\{v$, $v_S^i$, $m_{h_1}$, $m_{h_2}$, $m_{h_3}$, $\alpha_1$, $\alpha_2\}$ as independent inputs, fixing $v=246.22$ GeV and $m_{h_1}=125.2$ GeV~\cite{ParticleDataGroup:2024cfk}. 
Eliminating $v_S^r$ between the two constraints yields
\begin{align}
(\mathcal{M}_S^2)_{12}(\mathcal{M}_S^2)_{33}
-(\mathcal{M}_S^2)_{13}(\mathcal{M}_S^2)_{32}=0.
\end{align}
Since the mass-matrix elements are expressed as
$(\mathcal{M}_S^2)_{ij}=\sum_k O_{ik}O_{jk}m_{h_k}^2$,
this constraint depends only on the remaining mixing angle $\alpha_3$ once the three scalar masses and $\alpha_1,\alpha_2$ are specified. We first solve this equation for $\alpha_3$ and then determine $v_S^r$ from $v_S^r=(\mathcal{M}_S^2)_{12}v_S^i/(\mathcal{M}_S^2)_{13}$.
The scalar potential parameters are subsequently reconstructed using the relations given above.

As discussed in Sec.~\ref{sec:DWbias}, a small explicit CPV term, either $a_1^i$ or $b_1^i$, is required to destabilize the CPDW. We subsequently introduce a small nonzero $a_1^i$ to generate the vacuum-energy bias. Since the bias is extremely small in the parameter region of interest, its effects on the VEVs, scalar masses, and mixing angles can be safely neglected.

The Higgs couplings to gauge bosons and fermions are, respectively, defined as
\begin{align}
\mathcal{L}_{h_iVV} &=\frac{1}{v}\sum_{i=1}^3g_{h_iVV}^{}h_i(m_Z^2Z_\mu Z^\mu+2m_W^2W_\mu^+W^{-\mu}), \\
\mathcal{L}_{h_i\bar{f}f} 
&= -\sum_{i=1}^3h_i\bar{f}\left(g_{h_i\bar{f}f}^S+i\gamma_5g_{h_i\bar{f}f}^P\right)f,
\label{hiff}
\end{align}
where $g_{h_iVV}^{} = O_{1i}$. and 
\begin{align}
g_{h_i\bar{f}f}^S
& = \frac{m_f}{v}O_{1i}+\frac{v}{2\Lambda}(c_{f}^rO_{2i}-c_{f}^iO_{3i}),\label{gS}\\
g_{h_i\bar{f}f}^P
& =\frac{v}{2\Lambda}(c_{f}^iO_{2i}+c_{f}^rO_{3i}),]\label{gP}
\end{align}
with the definition $c_f = c_f^r+ic_f^i$. 
Note that $g_{h_i\bar{f}f}^P$ can be nonzero even when $c_f^i=0$. 
In contrast, in the absence of dimension-five Yukawa interactions, $g_{h_i\bar{f}f}^P$ vanishes, implying that CPV phases in the singlet scalar sector alone cannot induce CPV fermion couplings.

\section{CP domain walls}\label{sec:cpdw}
In the real basis, the $S$ and $S^2$ terms in the scalar potential $V_0$ become
\begin{align}
V_0\ni a_1^r(S+S^*)+\frac{b_1^r}{4}(S^2+S^{2*}) = 2a_1^r\mathrm{Re}S+\frac{b_1^r}{2}\mathrm{Re}S^2.
\end{align}
As a result, the whole $V_0$ is invariant under $S\to S^*$ or $\mathbb{Z}_2$ symmetry $\mathrm{Im}\,S\to -\mathrm{Im}\,S$.
Once $S$ develops the complex VEV $v_S$, $V_0$ has the two degenerate minima $V_0(v, v_{S}) = V_0(v, v_{S}^*)$.
These two degenerate vacua correspond to CP-conjugate minima and therefore lead to the formation of CPDWs separating regions with opposite CP phases.

Now, we derive the analytic expression of $v_S^i$. In the real basis, Eqs.~(\ref{tad_s}) and (\ref{tad_chi}) are reduced to
\begin{align}
\left\langle \frac{\partial V_0}{\partial s}\right\rangle & = 
v_S^r
\left[
\frac{b_2}{2}+\frac{\delta_2}{4}v^2+\frac{d_2}{4}|v_S|^2+\frac{b_1^r}{2}
\right]+\sqrt{2}a_1^r = 0, \label{vSr_real} \\
\left\langle \frac{\partial V_0}{\partial \chi}\right\rangle & = 
v_S^i
\left[
\frac{b_2}{2}+\frac{\delta_2}{4}v^2+\frac{d_2}{4}|v_S|^2-\frac{b_1^r}{2}
\right] = 0.\label{vSi_real}
\end{align}
Note that if $b_2-b_1^r+\delta_2v^2/2>0$, only a solution $v_S^i=0$ exists, while $v_S^r\neq0$ as long as $a_1^r\neq0$.
In the case of $v\neq0$, $v_S^r\neq0$, and $v_S^i\neq0$, $v_S^i$ is given by
\begin{align}
v_S^i = \pm \sqrt{-v_S^{r2}+\frac{2\lambda}{\delta_2^2-\lambda d_2}
\left[-\frac{\delta_2m^2}{\lambda}+b_2+\frac{\sqrt{2}a_1^r}{v_S^r}\right]}.
\end{align}
These two degenerate minima give rise to CPDW. 

Eq.~(\ref{scpv_J}) in the real basis takes the form
\begin{align}
\sin\theta_{v_S}|v_S|\Big[\sqrt{2}|a_1^r|\pm|b_1^r||v_S|\cos\theta_{v_S}\Big] = 0.
\end{align}
where the lower sign is for $\theta_{a_1}=\pi$.
The solution of $\sin\theta_{v_S} =0$ corresponds to the CP conserving case, while the CPV solution is
\begin{align}
\cos\theta_{v_S} = \mp \frac{\sqrt{2}|a_1^r|}{|b_1^r||v_S|}.
\end{align}
The solution exists if $|\cos\theta_{v_S}|\le 1$:
\begin{align}
|a_1^r| \le \frac{|b_1^r||v_S|}{\sqrt{2}}.
\label{a1r_upper}
\end{align}
Using  Eqs.~\eqref{vSr_real} and \eqref{vSi_real}, $b_1^r$ can be eliminated, and the condition (\ref{a1r_upper}) becomes $|v_S^r|\le |v_S|$, and thus $|\cos\theta_{v_S}|\le 1$ for any case in the real basis.

Let us obtain the configuration of CPDW. We parameterize the classical scalar fields as 
\begin{align}
\langle H(z)\rangle=\frac{1}{\sqrt{2}}
\left(
	\begin{array}{c}
		0 \\
		\phi(z)
	\end{array}
\right),\quad
\langle S(z)\rangle=\frac{1}{\sqrt{2}}\big( \phi_S^r(z)+i\phi_S^i(z)\big).\label{dw_fields}
\end{align}
The energy density of the DW is given by
\begin{align}
\mathcal{E}_{\text{DW}} =
	\frac{1}{2}(\partial_z \phi)^2+\frac{1}{2}(\partial_z \phi_S^r)^2+\frac{1}{2}(\partial_z \phi_S^i)^2+V(\phi, \phi_S^r, \phi_S^i),
\end{align}
where $V(\phi, \phi_S^r, \phi_S^i)=V_0(\phi, \phi_S^r, \phi_S^i)-V_0(v, v_S^r, v_S^i)$ with
\begin{align}
V_0(\phi, \phi_S^r, \phi_S^i) & = \frac{m^2}{4}\phi^2+\frac{\lambda}{16}\phi^4+\frac{\delta_2}{8}\phi^2(\phi_S^{r2}+\phi_S^{i2})+\frac{d_2}{16}(\phi_S^{r2}+\phi_S^{i2})^2 \nonumber\\
&\quad +\sqrt{2}a_1^r\phi_S^r+\frac{b_1^r}{4}(\phi_S^{r2}-\phi_S^{i2})
+\frac{b_2}{4}(\phi_S^{r2}+\phi_S^{i2}).
\end{align}
The equations of motion for the three fields $\Phi=\{\phi, \phi_S^r, \phi_S^i\}$ can be collectively written as
\begin{align}
\frac{d^2\Phi}{dz^2}-\frac{\partial V}{\partial \Phi}&=0, 
\label{eom}
\end{align}
with the boundary conditions
\begin{align}
\lim_{z\to\pm\infty}\phi(z)&=v,\quad \lim_{z\to\pm\infty}\phi_S^r(z)=v_S^r,\quad \lim_{z\to\pm\infty}\phi_S^i(z)=\pm v_S^i.
\label{bc}
\end{align}
The DW tension can be obtained by
\begin{align}
\sigma_{\text{DW}} = \int_{-\infty}^\infty dz~\mathcal{E}_{\text{DW}}.
\end{align}

In the small-mixing limit $\alpha_{1,2,3} \ll 1$, the DW profile is well approximated by the single-field solution of the $\phi^4$ theory. 
Its profile has the well-known form
\begin{align}
\phi_S^i(z) = v_S^i\tanh\left(\frac{d_2}{8}v_S^i(z-z_0)\right),
\end{align}
where $z_0$ is an arbitrary constant, which is the consequence of the translational symmetry. 
The DW tension is expressed as
\begin{align}
\sigma_\text{DW}&= \frac{2}{3}\sqrt{\frac{d_2}{2}}v_S^{i3}\simeq \frac{2}{3}m_{h_3}v_S^{i2},
\label{sigmaDW}
\end{align}
where $m_{h_3}^2\simeq d_2v_S^{i2}/2$.
Although Eq.~(\ref{sigmaDW}) provides a good approximation for the DW tension, we numerically solve the equations of motion for the DW, Eq.~(\ref{eom}), subject to the boundary conditions in Eq.~(\ref{bc}).

%
%
\subsection{The bias term for the DW collapses}\label{sec:DWbias}
In the real basis, the effective potential satisfies
\begin{align}
V_{\rm eff}(v,v_S)=V_{\rm eff}(v,v_S^*) ,
\end{align}
implying $\Delta V=0$, where
\begin{align}
\Delta V
= \big|V_\mathrm{eff}(v, v_S)-V_\mathrm{eff}(v, v_S^*)\big|
= \big|V_\mathrm{eff}(v, v_S^r, v_S^i)
     -V_\mathrm{eff}(v, v_S^r, -v_S^i)\big| .
\end{align}
To lift the degeneracy between the CP-conjugate vacua and avoid a
DW-dominated universe, we introduce soft $\mathbb{Z}_2$-breaking terms
in the scalar potential:
\begin{align}
V_{\ztwosla_2}(\varphi,\varphi_S^r, \varphi_S^i)
= -\sqrt{2}a_1^i\varphi_S^i
  -\frac{b_1^i}{2}\varphi_S^r\varphi_S^i .
\end{align}
These terms generate an energy difference between the two vacua,
\begin{align}
\Delta V
= 2|v_S^i|
  \left|\sqrt{2}a_1^i+\frac{b_1^i}{2}v_S^r\right| .
\end{align}
Note that there exists a parameter space where $\Delta V=0$ but
$I_1\neq0$, namely,
\begin{align}
a_1^i = -\frac{b_1^iv_S^r}{2\sqrt{2}} .
\end{align}
Hence, CPV is necessary but not sufficient to generate a DW bias through $V_{\ztwosla_2}$.
Even if the vacuum energy splitting $\Delta V$ is fixed, the underlying CPV structure is not uniquely determined.
Different distributions of the same $\Delta V$ among the CPV parameters lead to distinct CPV invariants and hence to different physical CPV effects. For example, we may consider the two limiting cases:
\begin{itemize}
\item[(i)] $a_1^i\neq 0,\quad b_1^i=0,
\quad \Delta V = 2\sqrt{2}|a_1^iv_S^i|$,
\item[(ii)] $a_1^i= 0,\quad b_1^i\neq0,
\quad \Delta V = |b_1^iv_S^rv_S^i|$.
\end{itemize}
For a fixed $\Delta V$, the magnitude of $a_1^i$ or $b_1^i$ is determined in each case, assuming that $v_S$ is given.
Although $\Delta V$ is a physical quantity and therefore invariant under a field redefinition $S=e^{i\theta_S}S'$, its explicit expression in terms of $a_1^i$, $b_1^i$, and $(v_S^r, v_S^i)$ depends on that convention.
Under such a redefinition, the real and imaginary components of both the couplings and the VEV are mixed, while the resulting
$\Delta V$ remains unchanged.
Therefore, classifications of the DW bias in terms of individual contributions proportional to $a_1^i$ or $b_1^i$ are phase-convention dependent, even though the total bias $\Delta V$ is basis independent and physical.
In our work, we adopt case (i), while case (ii) is adopted in Ref.~\cite{Chen:2020wvu}.
In any case, although the bias term is relevant for the DW dynamics, it is found to be numerically irrelevant for the EDM calculation as their magnitudes are tiny in our case.

The upper bound on $\Delta V$ follows from the condition for the existence of DWs~\cite{Saikawa:2017hiv}:
\begin{align}
\frac{\Delta V}{V_\mathrm{bar}}<0.795,\label{DelV_upper}
\end{align}
where $V_\mathrm{bar}$ is the height of the potential barrier separating the two CP-conjugate vacua. 
We evaluate this quantity using the unbiased CP-symmetric potential, setting $a_1^i=b_1^i=0$, 
and define
\begin{align}
V_\mathrm{bar}\equiv V_0(\Phi_\mathrm{top})-V_0(v,v_S^r,v_S^i),
\end{align}
where $\Phi_\mathrm{top}$ denotes the field configuration at the top of the barrier along the DW trajectory. 
This definition is invariant under a constant shift of the scalar potential.
The condition \eqref{DelV_upper} yields
\begin{align}
|a_1^i|<\frac{0.795}{2\sqrt{2}} \frac{V_\mathrm{bar}}{|v_S^i|} \simeq 0.281\frac{V_\mathrm{bar}}{|v_S^i|}.
\end{align}
In the small-mixing limit, $\Phi_{\rm top}\simeq(v,v_S^r,0)$, $V_\mathrm{bar}$ is approximated as
\begin{align}
V_\mathrm{bar} \simeq \frac{d_2}{16}v_S^{i4} \simeq \frac{m_{h_3}^4}{4d_2}.
\end{align}
The resulting upper bound on $|a_1^i|$ is much weaker than that imposed by the requirement of detectable GWs, as discussed in the next section.

We now comment on a possible bias arising from the dimension-five Yukawa
interactions.
The field-dependent top mass is
\begin{align}
\bar{m}_f^2
= \frac{|y_t(\varphi_S)|^2}{2}\varphi^2 ,
\end{align}
where
$y_t(\varphi_S)
= Y_t + c_t\varphi_S/(\sqrt{2}\Lambda)$
with
$Y_t=(V_L^{u\dagger}Y^uV_R^u)_{33}$.
For real $Y_t$ and $c_t$, the $\varphi_S^i$ dependence drops out:
\begin{align}
|y_t(\varphi_S)|^2
= Y_t^2
+\frac{c_t^2}{2\Lambda^2}|\varphi_S|^2
+\frac{\sqrt{2}}{\Lambda}Y_tc_t\varphi_S^r .
\end{align}
Since the fermionic one-loop contribution to the effective potential depends on the background only through the field-dependent
mass squared, $V^{(1)}_t(\bar m_t^2)$, no vacuum-energy splitting between $v_{S,+}$ and $v_{S,-}$ is induced at the one-loop level in this CP-symmetric limit.
Generally, the CP symmetry (existence of the real basis) and the DW bias are closely related. 
By definition, the effective potential is obtained from the effective action evaluated on constant background fields.
In a CP-symmetric limit, the effective action satisfies $\Gamma[S]=\Gamma[S^*]$,
and therefore $V_{\rm eff}(S)=V_{\rm eff}(S^*)$ to all loop orders, implying $\Delta V=0$.
In the present study, we restrict to real dimension-five Yukawa couplings, $c_t, c_e\in\mathbb{R}$, as a simplifying assumption.

\section{Gravitational waves from CP domain wall collapse}\label{sec:gw}
As seen in Eq.~(\ref{sigmaDW}), the DW tension is proportional to $v_S^{i3}$ or  $m_{h_3}v_S^{i2}$ in the small mixing angle limit.
Then, we introduce the dimensionless quantity $\hat{\sigma}_\mathrm{DW}$ as
\begin{align}
\sigma_{\text{DW}} = m_{h_3}v_S^{i2}\hat{\sigma}_{\text{DW}}.
\end{align}
We assume a conventional scaling DW network formed after a thermal phase transition~\cite{Saikawa:2017hiv}, and that the network enters the standard scaling regime before annihilation. 
The annihilation occurs when the pressure from the vacuum bias overcomes the wall tension. 
Accordingly, the annihilation time $t_{\rm ann}$ is determined by balancing the wall tension against the bias-induced pressure:
\begin{equation}
t_{\rm ann} = C_{\rm ann} \mathcal{A} \frac{\sigma_{\rm DW}}{\Delta V},
\end{equation}
where $C_{\rm ann} \simeq 2\text{--}5$~\cite{Saikawa:2017hiv} and $\mathcal{A} \simeq 0.8 \pm 0.1$~\cite{Hiramatsu:2013qaa}. 
For definiteness, we use $C_{\rm ann}=2$ and $\mathcal{A}=0.8$ in the following discussion.

Requiring that DWs disappear before the big-bang-Nucleosythesis (BBN) era, $t_{\rm ann} < t_{\rm BBN} \equiv 0.01~\text{s}$, we obtain a lower bound on $|a_1^i|$ as
\begin{align}
|a_1^i| > 2.3\times 10^{-15}~\text{GeV}^3
\left(\frac{m_{h_3}}{1~\text{TeV}}\right)
\left(\frac{v_S^i}{100~\text{TeV}}\right)
C_{\rm ann}\mathcal{A}\hat{\sigma}_{\rm DW},
\label{a1_low_BBN}
\end{align}
where $m_{h_3}=1~\text{TeV}$ and $v_S^i = 100~\text{TeV}$ are benchmark values relevant for the GW observability, as discussed below.

Assuming that DWs are annihilated during the radiation-dominated era, the temperature at $t = t_{\rm ann}$ is given by
\begin{align}
T_{\rm ann} 
&= 5.7~\text{MeV}
\left(\frac{g_*(T_{\rm ann})}{10}\right)^{-1/4} 
\left(\frac{|a_1^i|}{10^{-15}~\text{GeV}^3}\right)^{1/2}
\left(\frac{1~\text{TeV}}{m_{h_3}}\right)^{1/2}
\left(\frac{100~\text{TeV}}{v_S^i}\right)^{1/2} \nonumber \\
&\quad \times 
\left(C_{\rm ann}\mathcal{A}\hat{\sigma}_{\rm DW}\right)^{-1/2},
\end{align}
where $g_*(T_{\rm ann})$ denotes the effective number of relativistic degrees of freedom at temperature $T_{\rm ann}$. 
In the range $1~\mathrm{MeV} < T_{\rm ann} \ll 100~\mathrm{MeV}$, we have $g_*(T_{\rm ann}) = 10.75$~\cite{Kolb:1990vq}.

We also require that DWs not come to dominate the energy density of the Universe. 
If the Universe is initially radiation dominated, the time at which DWs would start to dominate can be estimated as
$t_{\text{dom}}=3m_{\text{pl}}^2/(32\pi\mathcal{A}\sigma_{\text{DW}})$~\cite{Saikawa:2017hiv},
where $m_{\text{pl}}=1.22\times 10^{19}$ GeV.
Requiring the annihilation of DWs to occur before this epoch, $t_{\text{ann}}<t_{\text{dom}}$, leads to
\begin{align}
|a_1^i|>8.0\times 10^{-18}~\text{GeV}^3
\left(\frac{m_{h_3}}{1~\text{TeV}}\right)^2
\left(\frac{v_S^i}{100~\text{TeV}}\right)^3
C_{\text{ann}}\mathcal{A}^2\hat{\sigma}_{\text{DW}}^2.
\label{a1_low_DW}
\end{align}
For the particular point $m_{h_3}=1$ TeV and $v_S^i = 100$ TeV, this bound is weaker than that obtained in Eq.~(\ref{a1_low_BBN}).
Nevertheless, since the bound scales with higher powers of $m_{h_3}$ and $v_S^i$, it may become more restrictive in regions where these parameters take larger values.

In this work, we use the formulas given in Ref.~\cite{Saikawa:2017hiv} to estimate the GW signature induced by the DW collapses.
The GW spectrum at peak frequency is given by
\begin{align}
f_{\text{peak}} &= 1.1\times 10^{-9}~\text{Hz}\left(\frac{g_*(T_{\text{ann}})}{10} \right)^{1/2}
\left(\frac{g_{*s}(T_{\text{ann}})}{10}\right)^{-1/3}\left(\frac{T_{\text{ann}}}{10~\text{MeV}} \right),\\
\Omega_{\text{GW}}(f_{\text{peak}})h^2 
&= 7.2\times 10^{-10}~\tilde{\epsilon}_{\text{GW}}\mathcal{A}^2
\left(\frac{g_{*s}(T_{\text{ann}})}{10}\right)^{-4/3}\left(\frac{T_{\text{ann}}}{10~\text{MeV}} \right)^{-4} \nonumber \\
&\hspace{3cm}\times \hat{\sigma}_{\text{DW}}^2
\left(\frac{m_{h_3}}{1~\text{TeV}}\right)^2\left(\frac{v_S^i}{100~\text{TeV}}\right)^4.
\end{align}
where $h=0.67$~\cite{ParticleDataGroup:2024cfk} denotes the reduced Hubble parameter, $\tilde{\epsilon}_{\text{GW}}=0.7\pm 0.4$, and $g_{*s}(T_\mathrm{ann})=10.75$ in our case. 
For an arbitrary frequency $f$, it is found that~\cite{Caprini:2009fx,Hiramatsu:2013qaa}
\begin{align}
\Omega_\mathrm{GW}(f)h^2 =\Omega_\mathrm{GW}(f_\mathrm{peak})h^2
\left[
\left(\frac{f}{f_{\text{peak}}}\right)^3\theta(f_\mathrm{peak}-f) 
+\left(\frac{f_{\text{peak}}}{f}\right)\theta(f-f_\mathrm{peak})
\right],\label{OmegGWh2}
\end{align}
where $\theta(f)$ denotes the step function. 

The signal-to-noise ratio (SNR) is given by~\cite{Breitbach:2018ddu,Schmitz:2020syl}
\begin{align}
\text{SNR} = \sqrt{n_\mathrm{det}\, t_{\text{obs}}\int_{f_{\text{min}}}^{f_{\text{max}}} df \left(\frac{\Omega_{\text{GW}}(f)h^2}{\Omega_{\text{exp}}(f)h^2} \right)^2},
\label{snr}
\end{align}
where $n_\mathrm{det}=1$ for autocorrelated detectors and $n_\mathrm{det}=2$ for cross-correlated detectors. 
We take $n_\mathrm{det}=1$ for THEIA and $n_\mathrm{det}=2$ for SKA. 
Here, $t_\mathrm{obs}$ denotes the observation time. 
In the present analysis, we scan the parameter space assuming $t_\mathrm{obs}=20~\mathrm{years}$.

For SKA, we use the noise function~\cite{Breitbach:2018ddu}
\begin{align}
\Omega_\mathrm{exp}(f) = \sqrt{\frac{2}{N_p(N_p-1)}}\frac{64\sqrt{3}\pi^4\sigma^2\,\delta t}{H_0^2}f^5,
\end{align}
where $N_p$ is the number of pulsars, $\delta t$ is the observation cadence, $\sigma$ is the timing uncertainty, and $H_0$ is the Hubble constant, parametrized as $H_0 = h\times 100\,\mathrm{km}\,\mathrm{s}^{-1}\,\mathrm{Mpc}^{-1}$.
We adopt $N_p = 50$, $\delta t = 7~\mathrm{days}$, and $\sigma = 100\,\mathrm{ns}$~\cite{Janssen:2014dka}. 
The frequency range is taken as $f_\mathrm{min}=t_\mathrm{obs}^{-1}$ and $f_\mathrm{max} = \delta t^{-1}$.

For THEIA, the noise function is given by~\cite{Garcia-Bellido:2021zgu,Roshan:2024qnv}
\begin{align}
\Omega_\mathrm{exp}(f) = \frac{2\pi^2}{3H_0^2}f^2h_\mathrm{GW}^2,
\end{align}
for $f > t_\mathrm{obs}^{-1}$, where $h_\mathrm{GW}$ denotes the strain sensitivity. 
We assume $h_\mathrm{GW} = 1.6\times 10^{-16}\,(1\,\mathrm{year}/t_\mathrm{obs})$ and take $f_\mathrm{max}=10^{-6}~\mathrm{Hz}$ when evaluating Eq.~(\ref{snr}). 

As a reference detectability criterion, we adopt $\mathrm{SNR}=20$ for both SKA and THEIA, following one of the benchmark choices considered in the literature~\cite{Chen:2020wvu}.
Qualitative conclusions remain unchanged for alternative choices, e.g., $\mathrm{SNR}=10$.
The detectability requirement of the GW signal places an upper limit on $|a_1^i|$, which follows from the scaling relation
$\Omega_\mathrm{GW}\propto T_{\rm ann}^{-4}\propto |a_1^i|^{-2}$.
The resulting upper bound must be consistent with the lower bound on $|a_1^i|$ obtained from either Eq.~(\ref{a1_low_BBN}) or Eq.~(\ref{a1_low_DW}).

Before closing, we comment on the initial-condition dependence for DWs.
The GW estimate presented above is based on the conventional analytic description of a DW network, which effectively assumes a standard thermal (white-noise) initial condition arising from a Kibble-type formation after a phase transition~\cite{Saikawa:2017hiv}.
In this case, the annihilation time is conventionally estimated by balancing the wall tension against the pressure induced by the vacuum bias, leading to the estimate of $t_{\rm ann}$ used in our analysis.
However, the DW dynamics can depend on the origin of the initial fluctuations.
If the wall-forming scalar remained light during inflation, the network may instead be characterized by scale-invariant correlations, potentially leading to a longer lifetime and lower GW peak frequency and amplitude compared to the conventional thermal case~\cite{Kitajima:2023kzu}.
Since the cosmological history of the singlet sector is not specified, our result should be regarded as a baseline estimate for the conventional thermal scenario.

For completeness, we comment on recent developments in numerical simulations of GWs from DW networks.
The benchmark values adopted in this work, such as $A = 0.8$ and $\tilde{\epsilon}_{\mathrm{GW}} = 0.7$, follow the conventional estimates summarized in Ref.~\cite{Saikawa:2017hiv}.
Recent lattice simulations have reported $A \simeq 0.8\text{--}0.9$, including $A = 0.78 \pm 0.03$~\cite{Cyr:2025nzf} and $A = 0.87 \pm 0.05$~\cite{Blasi:2025tmn}, consistent with our benchmark choice, while possible sensitivities to infrared initial conditions and lattice artifacts have also been discussed~\cite{Dankovsky:2025pjg}.
Recent studies further suggest that the GW spectrum may receive sizable corrections from the collapse dynamics of DWs beyond the instantaneous annihilation approximation.
In particular, GW production may continue efficiently during the collapse stage, enhancing the GW amplitude and modifying the spectral shape near the peak frequency~\cite{Notari:2025kqq,Cyr:2025nzf}, while other analyses suggest earlier annihilation and a suppressed GW signal~\cite{Babichev:2025stm}.
These results indicate that the precise GW normalization and spectral shape from collapsing DWs still carry non-negligible theoretical uncertainties.
Since the main purpose of this work is to study the complementarity between GW observables and the electron EDM, rather than to provide a precision prediction of the GW spectrum itself, we employ the conventional parametrization as a representative benchmark throughout this work.

\section{Electron EDM}\label{sec:edm}

EDMs are highly sensitive probes of CPV and provide powerful tools for exploring physics beyond the SM.
In particular, recent years have seen remarkable progress in searches for the electron EDM.
The most stringent constraints have been obtained by molecular experiments, notably those performed by the ACME Collaboration~\cite{ACME:2018yjb} and the JILA Collaboration~\cite{Roussy:2022cmp}.
Currently, the strongest upper bound is set by the JILA Collaboration:
\begin{align}
|d_e| < 4.1\times 10^{-30}~e\,\mathrm{cm}\quad (90\%\,\mathrm{CL}).
\end{align}
This bound significantly constrains possible sources of CPV beyond the SM.

In this model, the dominant contributions to $d_e$ arise from the two-loop diagrams. For convenience, we parametrize $d_e$ as
\begin{align}
d_e = d_e^\mathrm{BZ} + d_e^\mathrm{kite},
\end{align}
where the first term originates from the Barr-Zee diagrams~\cite{Barr:1990vd,Ellis:2008zy,Cheung:2009fc,West:1993tk,Chang:2005ac,Ellis:2010xm,Cheung:2014oaa,Inoue:2014nva,Bowser-Chao:1997kjp,Abe:2013qla}, 
while the second arises from the kite diagrams~\cite{Altmannshofer:2020shb}. 
Although $d_e^\mathrm{kite}$ can dominate in certain regions of parameter space, as shown in Ref.~\cite{Altmannshofer:2020shb}, it is found to be subdominant in our setup.
In the following, we focus on the dominant Barr-Zee contributions to clarify the parametric dependence of $d_e$, while the kite contributions are fully included in the numerical analysis.

Depending on the particles running in the loop, we decompose their contributions into two parts
\begin{align}
d_e^\mathrm{BZ}  = d_e^{h \gamma}+d_e^{h Z}.
\label{de_BZ}
\end{align}
It is well known that $d_e^{h\gamma}$ dominates over $d_e^{hZ}$, as the latter is suppressed by the factor 
$(-1/4+\sin^2\theta_W)\simeq -0.02$, with $\theta_W$ denoting the weak-mixing angle~\cite{ParticleDataGroup:2024cfk}.

To illustrate the CPV dependence of $d_e^\mathrm{BZ}$, we keep $I_{1,2,3}\neq0$ in this section, while they are set to zero in the numerical analysis. Since the CPV phase dependence of $d_e^{hZ}$ is identical to that of $d_e^{h\gamma}$, we focus on $d_e^{h\gamma}$.

From Eqs.~(\ref{gS}), (\ref{gP}), (\ref{de_hgam_t}), and (\ref{de_hgam_W}), the top and $W$-loop contributions to the electron EDM are, respectively, given by 
\begin{align}
\frac{(d_e^{h\gamma})_t}{e} 
& = \frac{\alpha_{\text{em}}m_e}{12\pi^3\Lambda}\sum_{i=1}^3O_{1i}
\bigg[
	\frac{c_{e}^rO_{3i}+c_{e}^iO_{2i}}{m_e}f(\tau_{th_i})+\frac{c_{t}^rO_{3i}+c_{t}^iO_{2i}}{m_t}g(\tau_{th_i})
\bigg], \\
\frac{(d_{e}^{h\gamma})_W}{e}
& = -\frac{\alpha_\mathrm{em}}{64\pi^3\Lambda}\sum_{i=1}^3(c_{e}^rO_{3i}+c_{e}^iO_{2i})O_{1i}\mathcal{J}_W^\gamma(m_{h_i}), 
\end{align}
where $\tau_{th_i}=m_t^2/m_{h_i}^2$. 
Note that $(d_e^{h\gamma})_t$ depends on both the new Yukawa couplings $c_t$ and $c_e$, while $(d_e^{h\gamma})_W$ depends only on $c_e$.

Let us consider a case in which $|\alpha_{1,2,3}|\ll 1$. The mixing matrix is simplified to
\begin{align}
O & \simeq 
\begin{pmatrix}
1 & 0 & 0 \\
0 & 1 & -\alpha_3 \\
0 & \alpha_3 & 1
\end{pmatrix}
\begin{pmatrix}
1 & 0 & -\alpha_2  \\
0 & 1 & 0 \\
\alpha_2 & 0 & 1
\end{pmatrix}
\begin{pmatrix}
1 & -\alpha_1 & 0 \\
\alpha_1 & 1 & 0 \\
0 & 0 & 1
\end{pmatrix} 
\simeq  
\begin{pmatrix}
1 & -\alpha_1 & -\alpha_2 \\
\alpha_1 & 1 & -\alpha_3 \\
\alpha_2 & \alpha_3 & 1
\end{pmatrix},
\end{align}
where we keep the mixing angles up to $\mathcal{O}(\alpha_i)$ and neglect $\mathcal{O}(\alpha_i\alpha_j)$ or higher.
The mixing angles are approximately given by
\begin{align}
\alpha_1 \simeq \frac{1}{2}\frac{\delta_2vv_S^r}{\Delta m_{12}^2} , \quad
\alpha_2 \simeq \frac{1}{2} \frac{\delta_2vv_S^i}{\Delta m_{13}^2}, \quad
\alpha_3 \simeq \frac{1}{2}\frac{-b_1^i + d_2v_S^rv_S^i }{\Delta m_{23}^2},
\end{align}
where $\Delta m_{ij}^2 = m_{h_i}^2-m_{h_j}^2$.
Under this approximation, the electron EDM is simplified to
\begin{align}
\frac{(d_e^{h\gamma})_t}{e} 
& = \frac{\alpha_{\text{em}}m_e}{12\pi^3\Lambda}
\bigg[
	\frac{1}{m_e}\left(c_e^r\alpha_2f_{13}+c_e^i\alpha_1f_{12} \right)
	+\frac{1}{m_t}\left(c_t^r\alpha_2g_{13}+c_t^i\alpha_1g_{12} \right)
\bigg] \nonumber\\
& = \frac{\alpha_{\text{em}}m_e\delta_2 v}{24\pi^3\Lambda}
\bigg[
	\frac{1}{m_e}\left(\frac{c_e^rv_S^i}{\Delta m_{13}^2}f_{13}+\frac{c_e^iv_S^r}{\Delta m_{12}^2}f_{12} \right)
	+\frac{1}{m_t}\left(\frac{c_t^rv_S^i}{\Delta m_{13}^2}g_{13}+\frac{c_t^iv_S^r}{\Delta m_{12}^2}g_{12} \right)
\bigg],\label{de_an}
\end{align}
where $f_{ij}=f(\tau_{th_i})-f(\tau_{th_j})$ and $g_{ij}=g(\tau_{th_i})-g(\tau_{th_j})$.
In this small mixing angle limit, the electron EDM is insensitive to $\alpha_3$.
Furthermore, the $\alpha_1$-dependent terms drop for $c_t^i=c_e^i=0$.

In the region of our interest, $m_t\ll m_{h_2}, m_{h_3}$, the electron EDM is further approximated as
\begin{align}
\frac{(d_e^{h\gamma})_t}{e} 
& = -\frac{\alpha_{\text{em}}m_e\delta_2 v}{24\pi^3\Lambda}
\bigg[
	\frac{f(\tau_{th_1})}{m_e}\left(\frac{c_e^rv_S^i}{m_{h_3}^2}+\frac{c_e^iv_S^r}{m_{h_2}^2} \right)
	+\frac{g(\tau_{th_1})}{m_t}\left(\frac{c_t^rv_S^i}{m_{h_3}^2}+\frac{c_t^iv_S^r}{m_{h_2}^2}\right)
\bigg] \nonumber\\
&\simeq -\frac{\alpha_{\text{em}}m_e\delta_2 v}{24\pi^3\Lambda m_H^2}
\bigg[
	\frac{\mathrm{Im}(v_Sc_e)f(\tau_{th_1})}{m_e}+\frac{\mathrm{Im}(v_Sc_t)g(\tau_{th_1})}{m_t}
\bigg].
\label{de_an_lim}
\end{align}
where the second line is obtained by taking $m_{h_2}\simeq m_{h_3}\equiv m_H$.
We note that the electron EDM is expressed in terms of the CP-odd basis invariants $J_t$ and $J_e$ in the real $y_{t,e}$ basis.

Similarly, the $W$-loop contribution to the electron EDM in the approximation $m_{h_1}\ll m_{h_2}, m_{h_3}$ with the degenerate $h_2$ and $h_3$ is expressed as
\begin{align}
\frac{(d_e^{h\gamma})_W}{e}
&=-\frac{\alpha_\mathrm{em}}{64\pi^3\Lambda}
(c_e^r\alpha_2\mathcal{J}^\gamma_{13}+c_e^i\alpha_1\mathcal{J}^\gamma_{12})
\simeq 
\frac{\alpha_\mathrm{em}\delta_2v}{128\pi^3\Lambda m_H^2}\mathrm{Im}(v_Sc_e)\mathcal{J}_W^\gamma(m_{h_1}),
\end{align}
where $\mathcal{J}^\gamma_{ij} = \mathcal{J}_W^\gamma(m_{h_i})-\mathcal{J}_W^\gamma(m_{h_j})$.

As widely discussed in the literature~\cite{Fuyuto:2019svr,Nhi:2025iob}, an accidental cancellation may occur between $(d_e^{h\gamma})_t$ and $(d_e^{h\gamma})_W$.
For $c_e^i=c_t^i=0$, the condition for the cancellation between the top and $W$-loop contributions is
\begin{align}
\frac{(d_e^{h\gamma})_t}{e}+\frac{(d_e^{h\gamma})_W}{e}
= \frac{\alpha_\mathrm{em}\alpha_2}{12\pi^3\Lambda}
\left[
	c_e^rf_{13}+\frac{y_e}{y_t}c_t^rg_{13}-\frac{3}{16}c_e^r\mathcal{J}_{13}^\gamma
\right] = 0,\label{de_tW_hgam}
\end{align}
from which we obtain
\begin{align}
c_e^r = \mathcal{C}\times \frac{y_e}{y_t}c_t^r,
\end{align}
where $\mathcal{C} = g_{13}/(\frac{3}{16}\mathcal{J}_{13}^\gamma-f_{13})$, which is typically $\mathcal{O}(1)$ in magnitude.
For instance, $\mathcal{C}\simeq 0.7$ for $m_{h_3}=1$ TeV.
Therefore, the approximate relation $c_e^r/c_t^r\simeq y_e/y_t$ realizes the condition for the accidental cancellation. 
Since the required hierarchical structure of $c_{t,e}$ resembles the SM Yukawa hierarchy, this mechanism is referred to as the \textit{structured cancellation}~\cite{Fuyuto:2019svr}.
In contrast, the top and $W$-loop contributions become constructive when $c_e^r/c_t^r\simeq -y_e/y_t$.
This feature plays an important role in the numerical analysis presented in the next section.

From Eq.~(\ref{de_tW_hgam}), the electron EDM scales as $d_e \propto \alpha_2/\Lambda$ for fixed new Yukawa couplings $c_{t,e}$ and $m_{h_3}$. 
As discussed in the next section, the cutoff scale $\Lambda$ must be larger than the relevant mass scales such as $v_S^i$ and $m_{h_3}(>m_{h_2})$, implying $d_e \propto \alpha_2/(\mathrm{max}(v_S^i,m_{h_3}))$.
On the other hand, Eq.~(\ref{sigmaDW}) shows that realizing a sizable $\sigma_\mathrm{DW}$ requires large $v_S^i$ and/or $m_{h_3}$.
These observations indicate that the magnitude of $\alpha_2$ plays an important role in accessing the parameter region where both $d_e$ and $\sigma_\mathrm{DW}$ are not strongly suppressed.
In the next section, we quantify the maximal value of $\alpha_2$ allowed by theoretical and experimental constraints.

\section{Numerical analysis}\label{sec:results}
We explore the parameter space by imposing theoretical and experimental constraints.

\noindent
{\bf Vacuum stability:} For the scalar potential to be bounded from below, $\lambda>0$ and $d_2>0$ must hold, and additionally, $\delta_2 > -\sqrt{\lambda d_2}$ if $\delta_2<0$.

\noindent
{\bf Global minimum:}
We ensure that the chosen minimum is the global minimum by requiring $\Delta V=V_0(\varphi, \varphi_S^r, \varphi_S^i)-V_\mathrm{min}>0$, where
\begin{align}
\Delta V &= \frac{\lambda}{16}(\varphi^2-v^2)^2
+\frac{d_2}{16}\left[|\varphi_S|^2-|v_S|^2\right]^2
+\frac{\delta_2}{8}
\Big[
	(\varphi^2-v^2)(|\varphi_S|^2-|v_S|^2)
\Big] \nonumber\\
& -\frac{a_1^r}{\sqrt{2}v_S^r}(\varphi_S^r-v_S^r)^2
+\frac{a_1^i}{\sqrt{2}v_S^i}(\varphi_S^i-v_S^i)^2.
\end{align}
The first two terms are positive definite, while the remaining terms could be either positive or negative. 

\noindent
{\bf Perturbative unitarity:}
We use the expressions shown in Ref.~\cite{Chen:2020wvu}.
In addition to these perturbative unitarity bounds in the scalar sector, we require that the EFT expansion of the dimension-five Yukawa interactions remains under perturbative control. 
In our setup, we fix
\begin{align}
c_t^i = c_e^i = 0, \qquad
|c_t^r| = y_t, \qquad
|c_e^r| = y_e ,
\end{align}
so that the corrections induced by the dimension-five operators are suppressed by the ratio $v/\Lambda$.
In this study, we consider two schemes for cutoff scales:
\begin{itemize}
\item $\Lambda = c\times \max\!\left(v_S^i, m_{h_3}\right)$,
\item Fixed cutoff $\Lambda$.
\end{itemize}
For the former scheme, $c$ characterizes the hierarchy between the cutoff and the physical scales, which we assume $c=5\text{--}10$. When scanning the parameter space with varying $v_S^i$ or $m_{h_3}$, $\Lambda$ accordingly changes in this scheme. In the second scheme, however, $\Lambda$ is held fixed during the parameter scan.

\subsection{Current experimental constraints}
\noindent
As discussed, the current upper bound on the electron EDM is $|d_e|<4.1\times 10^{-30}~e\,\mathrm{cm}$.
Other than this, we consider the experimental constraints from colliders and cosmological observations, which are described below.

\noindent
{\bf LHC constraints:}
In our model, the top Yukawa coupling can be modified by the mixing among the three scalars and the dimension-five operator, as in Eq.~(\ref{hiff}). 
The CMS Collaboration places constraints on the Higgs couplings to the top quark,
defined as $\kappa_t \equiv v g_{h_1\bar{t}t}^S/m_t$ and 
$\tilde{\kappa}_t \equiv v g_{h_1\bar{t}t}^P/m_t$,
at $95\%$ CL, and on the CPV parameter $f_{\rm CP}^{tth}$ at $68\%$ CL~\cite{CMS:2022dbt}:
\begin{align}
0.86 \le \kappa_t \le 1.26,\quad
-1.07 \le \tilde{\kappa}_t \le 1.07,\quad  |f_{\rm CP}^{htt}| =
\frac{|\tilde{\kappa}_t|^2}{|\kappa_t|^2+|\tilde{\kappa}_t|^2}
< 0.55 .
\end{align}

The Higgs coupling to gauge bosons is rescaled by the doublet component of the scalar mass eigenstate, $\kappa_V \equiv g_{h_1VV} = O_{11}$. In the absence of additional non-SM decay modes, the inclusive Higgs signal strength approximately satisfies $\mu \simeq \kappa^2$.
The run~2 measurements give $\mu_\mathrm{ATLAS}=1.05\pm0.06$~\cite{ATLAS:2022vkf} and  $\mu_\mathrm{CMS}=1.002\pm0.057$~\cite{CMS:2022dwd}, which are mutually consistent. In the following analysis, we use the ATLAS result as a reference constraint.
Using the approximate relation $\mu \simeq \kappa^2$ together with the model condition $\kappa \le 1$, the ATLAS $95\%$ CL range is translated into $0.964 \lesssim \kappa \le 1$.
Note, however, that the ATLAS and CMS analyses are carried out assuming the SM Higgs production and decay structure.
Since the dimension-five top Yukawa interaction in our model modifies the Higgs production rates,
the above bound should be interpreted only as an approximate reference constraint on the doublet component of $h_1$.

Regarding the constraints on $m_{h_2}$ and $m_{h_3}$, since the heavier scalars are assumed to lie around the TeV or higher scales and their production rates are suppressed by the small doublet component, current direct searches for heavy scalars at the LHC do not provide significant additional constraints in the parameter region considered in this work.

\noindent
{\bf $S$, $T$, $U$ constraints:}
Electroweak precision constraints from the oblique parameters~\cite{Peskin:1990zt,*Peskin:1991sw} can be implemented in the same way as in the cxSM scalar sector.
Since the dimension-five top Yukawa interaction does not modify the electroweak gauge couplings at tree level, its effect on the oblique parameters is subleading, while the dominant contributions arise from the scalar mixing and the heavy-scalar masses $m_{h_{2,3}}$.
In practice, we therefore evaluate the scalar-sector contributions to $S$, $T$, and $U$ and require consistency with the current global electroweak fit. In the parameter region considered in this work, $m_{h_2},m_{h_3}\sim {\cal O}(1)$ TeV or higher and the observed Higgs
boson is SM-like with $O_{11}\simeq 1$. The heavier scalars are therefore mostly singletlike, so their contributions to the oblique parameters are mixing suppressed. As a result, the $S$, $T$, and $U$ constraints are typically weaker than the Higgs signal strength and EDM bounds in the parameter region of interest.

\noindent
{\bf Cosmological constraints:}
We also consider cosmological constraints. 
As discussed in Sec.~\ref{sec:gw}, DWs must disappear before the onset of BBN~\cite{Kolb:1990vq}. 
Although the measurement of $\Delta N_\mathrm{eff}$ could in principle constrain scenarios with sizable $\sigma_\mathrm{DW}$, it does not provide a useful bound in the parameter space considered below.
\subsection{Future prospects}\label{subsec:future}
Several ongoing and proposed molecular experiments aim to further improve the sensitivity of the electron EDM.
In particular, experiments using polar molecules and laser-cooled polyatomic molecules are expected to enhance the sensitivity by 1 to 2 orders of magnitude relative to the current limit (see, e.g., Ref.~\cite{Kozyryev:2017cwq}).
In addition, a proposal based on matrix-isolated polar molecules suggests that a statistical sensitivity at the level of
$10^{-34}\,e\,{\rm cm}$ may be achievable in an idealized experimental setup~\cite{Vutha:2018tsz}.
These future improvements will further extend the reach of EDM searches for new CPV sources.
For illustration, we show prospective sensitivities ranging from $10^{-33}$ to $10^{-31}~e\,\mathrm{cm}$.

As relevant GW observatories, we consider SKA~\cite{Janssen:2014dka} and THEIA~\cite{Theia:2017xtk}, 
and explore the parameter space by imposing $\mathrm{SNR}=20$ for each experiment.
As discussed in Sec.~\ref{sec:gw}, the GW spectrum in Eq.~(\ref{OmegGWh2}) is determined by the DW tension $\sigma_\mathrm{DW}$ and the bias term $\Delta V$.
In our analysis, $\sigma_\mathrm{DW}$ is computed from the model parameters, while $\Delta V$ is fixed to its minimal value consistent with cosmological constraints, in particular the BBN bound. 
This condition implies $\hat{\sigma}_\mathrm{DW}\ge 1.2\times 10^4$ for SKA and $\hat{\sigma}_\mathrm{DW}\ge 2.8\times 10^3$ for THEIA. 
We identify these regions as those where the GW signal is detectable.

Future Higgs factories, such as the ILC~\cite{Fujii:2017vwa}, FCC-ee~\cite{FCC:2018evy}, and CEPC~\cite{CEPCStudyGroup:2018ghi} will significantly improve the precision of Higgs coupling measurements.
For the 250\,GeV stage of the ILC with 2\,ab$^{-1}$, the expected precision is $\delta\kappa_V \simeq 0.38\%$, corresponding to the approximate $95\%$ CL range $0.992 \lesssim \kappa_V \lesssim 1.008$.
The FCC-ee is expected to further improve this precision to $\delta\kappa_V \sim 0.2\%$, leading to $0.996 \lesssim \kappa_V \lesssim 1.004$. Note that CEPC could also provide a similar precision.

\subsection{Results}
The free model parameters are $\{v_S^i,\, m_{h_2},\, m_{h_3},\, \alpha_1,\, \alpha_2; c_t^r, c_e^r, \Lambda; a_1^i \}$.
Throughout our numerical analysis, we fix $c_t^r = y_t$.
Our aim is to probe the parameter region where sizable CPV occurs while remaining consistent with current experimental constraints. 
Since $\kappa_V = O_{11} = c_1 c_2$, the mixing angles $\alpha_1$ and $\alpha_2$ cannot be too large in order to realize $\kappa_V \simeq 1$. 
On the other hand, as shown in Eq.~(\ref{de_an}), the $\alpha_2$-dependent terms play an important role in the electron EDM in the case of SCPV where $c_t^i = c_e^i = 0$. 
Motivated by this, we fix $\alpha_1 = {10^{-4}}^\circ$ and treat $\alpha_2$ as a free parameter.
We first present the results for $\Lambda=c\times\max(v_S^i,m_{h_3})$, followed by those for a fixed cutoff $\Lambda$.

As discussed in the previous two sections, the mass scale $m_{h_3}$ plays a key role, since $\Omega_\mathrm{GW}\propto m_{h_3}^2$, while $d_e\propto 1/m_{h_3}$ for $\Lambda = c\, m_{h_3}$. 
In contrast, $m_{h_2}$ has a smaller impact on these observables. 
However, a large mass splitting between $h_2$ and $h_3$ tends to violate perturbative unitarity. 
We therefore consider two representative benchmark points (BPs):
\begin{align}
\mathrm{BP1}:&\quad m_{h_2} = 0.9~\mathrm{TeV},\quad m_{h_3} = 1~\mathrm{TeV}, \\
\mathrm{BP2}:&\quad m_{h_2} = 9~\mathrm{TeV},\quad m_{h_3} = 10~\mathrm{TeV}.
\end{align}
These benchmarks illustrate the interplay between GW and EDM observables across different mass scales and serve as representative points. Comments on other parameter spaces will be made at the end of this section.

For each benchmark point, we consider $(c_t^r, c_e^r) = (y_t, 0)$ and $(c_t^r, c_e^r)=(y_t, -y_e)$, respectively, and explore the regions compatible with all the constraints and detectable GW signals by scanning $\alpha_2$ and $v_S^i$.
Note that  $c_e^r$ is chosen in order not to induce the cancellation in the electron EDM. 
We further categorize BPs into $\Lambda = 5 \mathrm{max}(v_S^i, m_{h_3})$ and $\Lambda = 10 \mathrm{max}(v_S^i, m_{h_3})$ cases, respectively.
The parameter $a_1^i$ is fixed by the bias term $\Delta V$ which is determined by the method described in Sec.~\ref{subsec:future}.

\begin{figure}[t]
\center
\includegraphics[width=7.5cm]{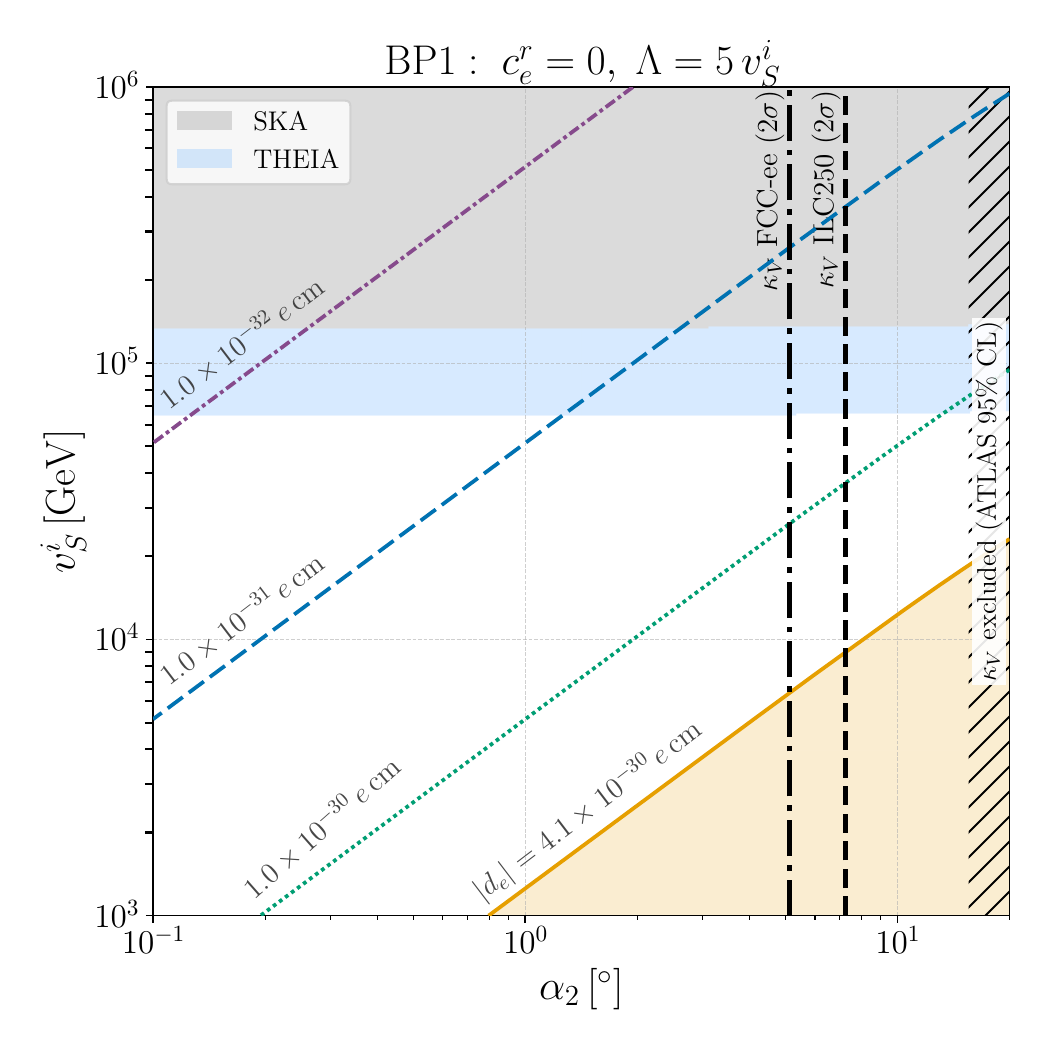}  \hspace{0.5cm}
\includegraphics[width=7.5cm]{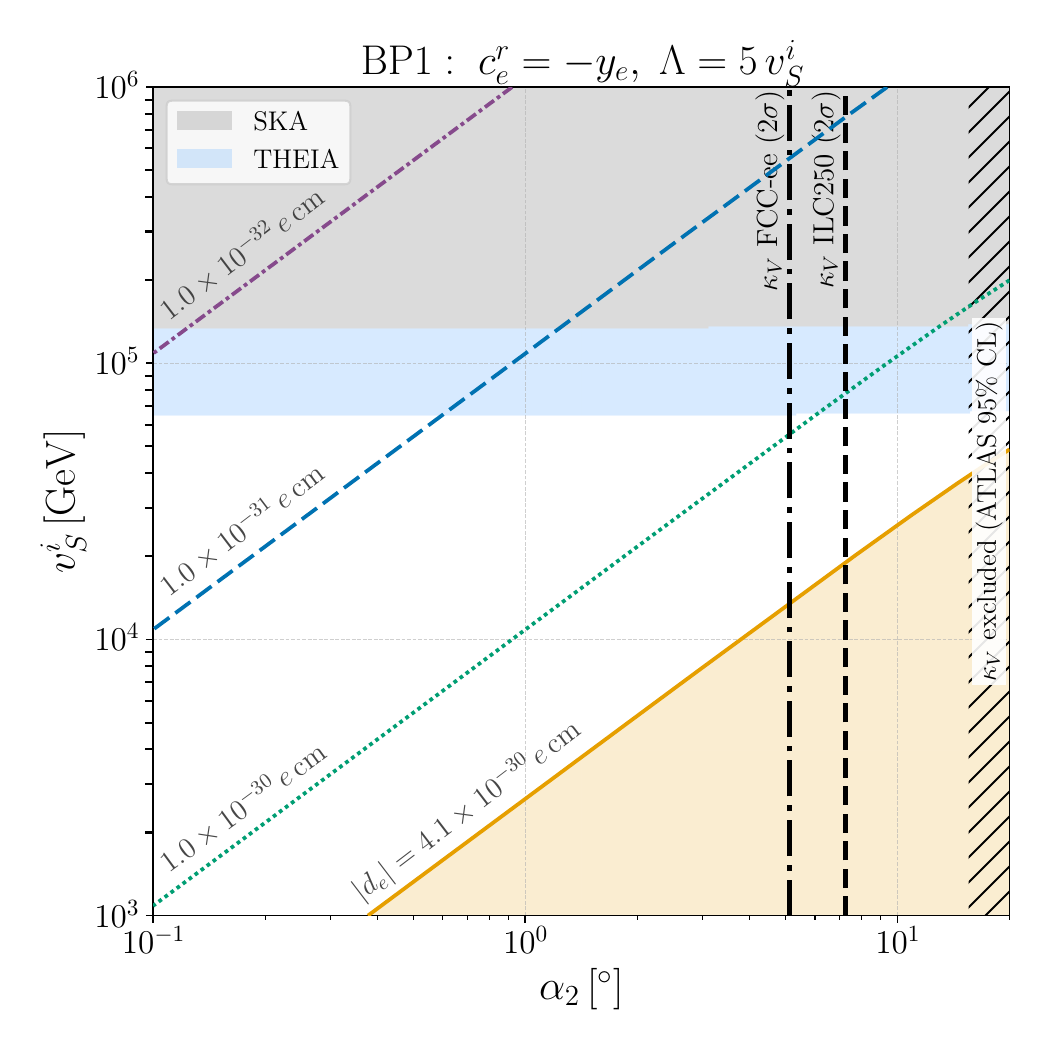} \\[1cm]
\includegraphics[width=7.5cm]{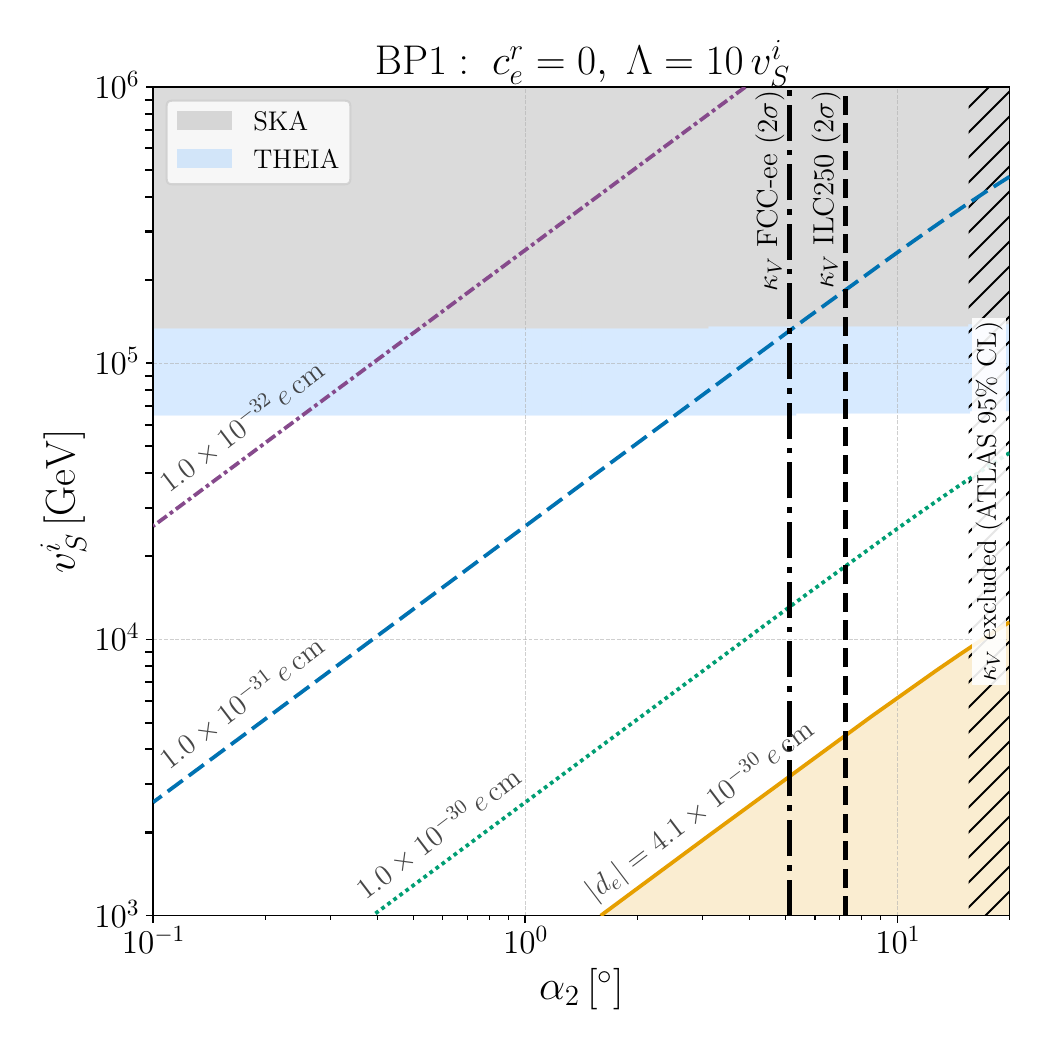} \hspace{0.5cm}
\includegraphics[width=7.5cm]{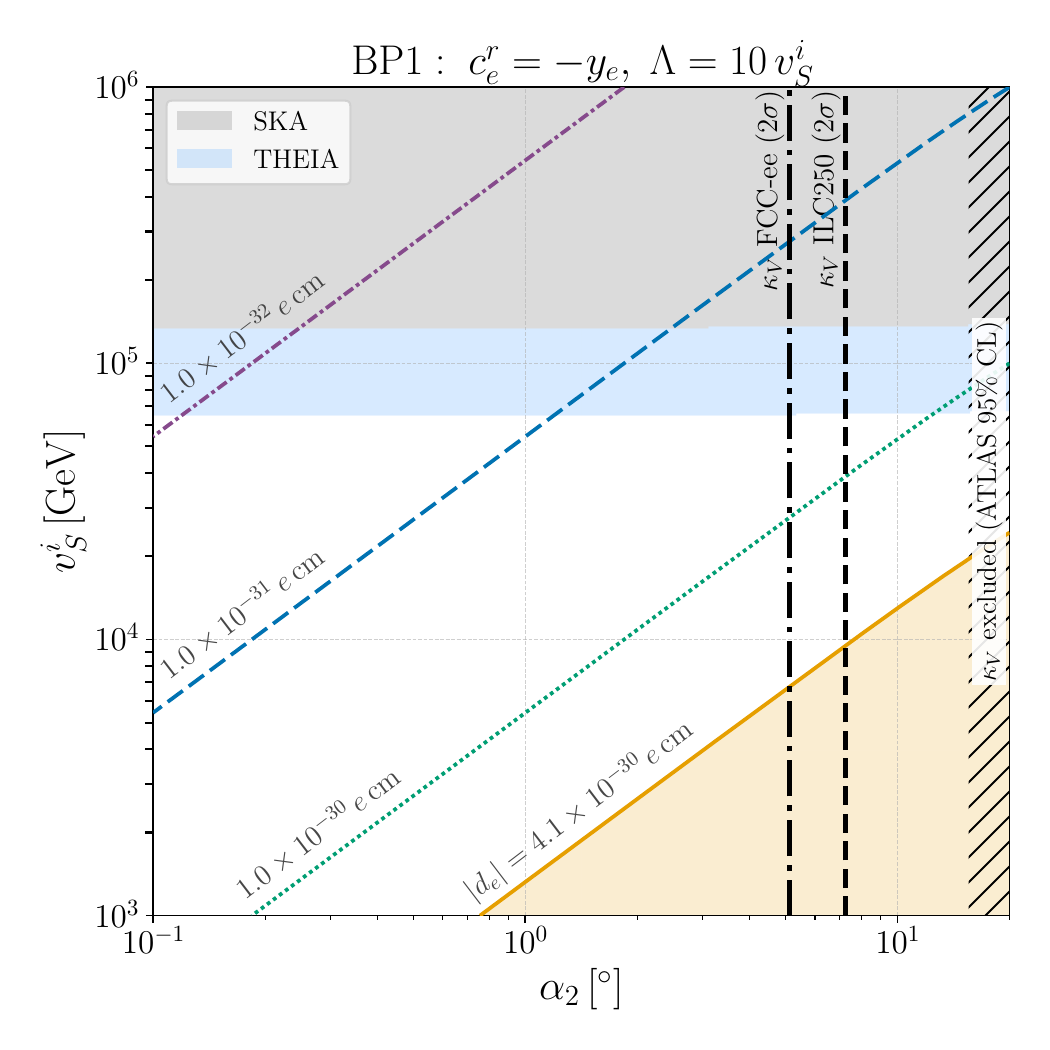}
\caption{
Contours of the electron EDM and the GW-detectable regions by SKA and THEIA are shown in the $(\alpha_2, v_S^i)$ plane for BP1 with $\Lambda = (5,10)v_S^i$. 
The orange shaded region is excluded by the current JILA bound, $|d_e| = 4.1\times10^{-30}\,e\,\mathrm{cm}$. 
The green dotted, dark-blue dashed, and purple dot-dashed curves correspond to $|d_e| = 1.0\times10^{-30}$, $1.0\times10^{-31}$, and $1.0\times10^{-32}\,e\,\mathrm{cm}$, respectively. 
The hatched region is excluded by the ATLAS measurement of $\kappa_V$ at $95\%$ CL, while the vertical dashed and dot-dashed lines indicate its projected $2\sigma$ sensitivities at the ILC and FCC-ee. 
The grey and light-blue shaded regions denote the GW-detectable parameter space for SKA and THEIA, respectively. 
The overlap between the EDM sensitivity and the GW-detectable regions highlights the complementarity of the two probes.
}
\label{fig:GW-de_alp2_vSi_bm1}
\end{figure}

In Fig.~\ref{fig:GW-de_alp2_vSi_bm1}, the contours of the electron EDM and the region where GW is detectable at SKA and THEIA are plotted in the $(\alpha_2, v_S^i)$ plane for BP1.
We take $(c_e^r, \Lambda) = (0, 5v_S^i)$ (upper left), $(-y_e, 5v_S^i)$ (upper right), $(0, 10v_S^i)$ (lower left), $(-y_e, 10v_S^i)$ (lower right), respectively.

$|d_e|$ in the shaded region in orange exceeds the current electron EDM upper bound by JILA $|d_e|= 4.1\times 10^{-30}~e~\mathrm{cm}$, while
each contour represents $|d_e|=1.0\times 10^{-30}~e~\mathrm{cm}$ (dotted, green), $|d_e|=1.0\times 10^{-31}~e~\mathrm{cm}$ (dashed, dark blue), and $|d_e|=1.0\times 10^{-32}~e~\mathrm{cm}$ (dot-dashed, purple), respectively.
The hatched region is excluded by the $\kappa_V$ measurement at ATLAS at 95\% CL.
The 2$\sigma$ sensitivities of $\kappa_V$ at ILC and FCC-ee are shown by the vertical dashed and dot-dashed lines, respectively.
The shaded grey and light blue regions denote the GW detectable by SKA and THEIA, respectively.

In the upper left panel with $(c_e^r, \Lambda) = (0, 5v_S^i)$, the current JILA bound probes the region with $v_S^i \lesssim 2\times10^4~\mathrm{GeV}$ for $\alpha_2 = \mathcal{O}(1)$, while a large portion of the parameter space remains unexplored. 
Future improvements in the electron EDM sensitivity down to the $10^{-32}$--$10^{-31}\,e\,\mathrm{cm}$ level would allow this region to be probed, including the SKA- and THEIA-sensitive region where $v_S^i \gtrsim 7\times 10^4~\mathrm{GeV}$. 
It should be noted that the apparent $v_S^i$ dependence of the electron EDM, $d_e \propto 1/v_S^i$, originates from our choice of the cutoff scale, $\Lambda = c\,\max(v_S^i, m_{h_3})$. 
We also note that future measurements of $\kappa_V$ provide a complementary probe of $\alpha_2$, to which the electron EDM is particularly sensitive.

The impact of the nonzero $c_e^i$ can be seen in the upper right panel, where $(c_e^r, \Lambda) = (-y_e, 5v_S^i)$.
As discussed in Sec.~\ref{sec:edm}, the sign of $c_e^r$ is crucial for the accidental cancellation in the electron EDM. For $c_e^r<0$, the contributions from the $W$-loop Barr-Zee and kite diagrams are constructive to the top-loop Barr-Zee diagrams. As a result, the JILA excluded region expands to $v_S^i \lesssim 5\times10^4$ GeV for the maximally allowed $\alpha_2$, and 
the future electron EDM experiments could cover a wider parameter space than the case of $c_e^i=0$. 
Note that the DW tension remains the same at the accuracy of our calculation, and thus the GW-detectable region is unchanged.  

The lower two plots clarify the dependence of the cutoff $\Lambda$ on the electron EDM by changing it from $ 5v_S^i$ to $ 10v_S^i$.
One can see that the electron EDM is reduced by a factor of 2. Nevertheless, the GW-detectable regions at SKA and THEIA are still covered by the future electron EDM experiments. 

\begin{figure}[t]
\center
\includegraphics[width=7.5cm]{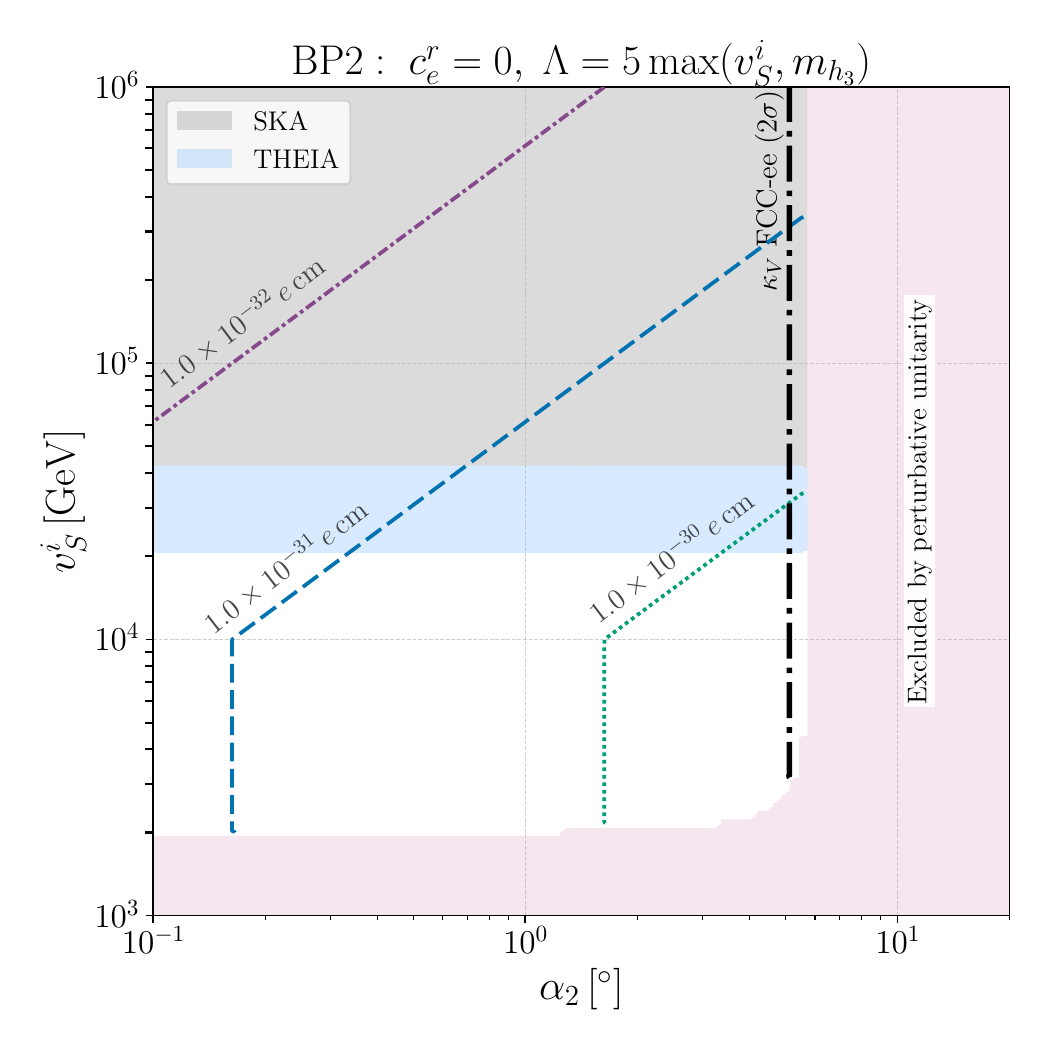}  \hspace{0.5cm}
\includegraphics[width=7.5cm]{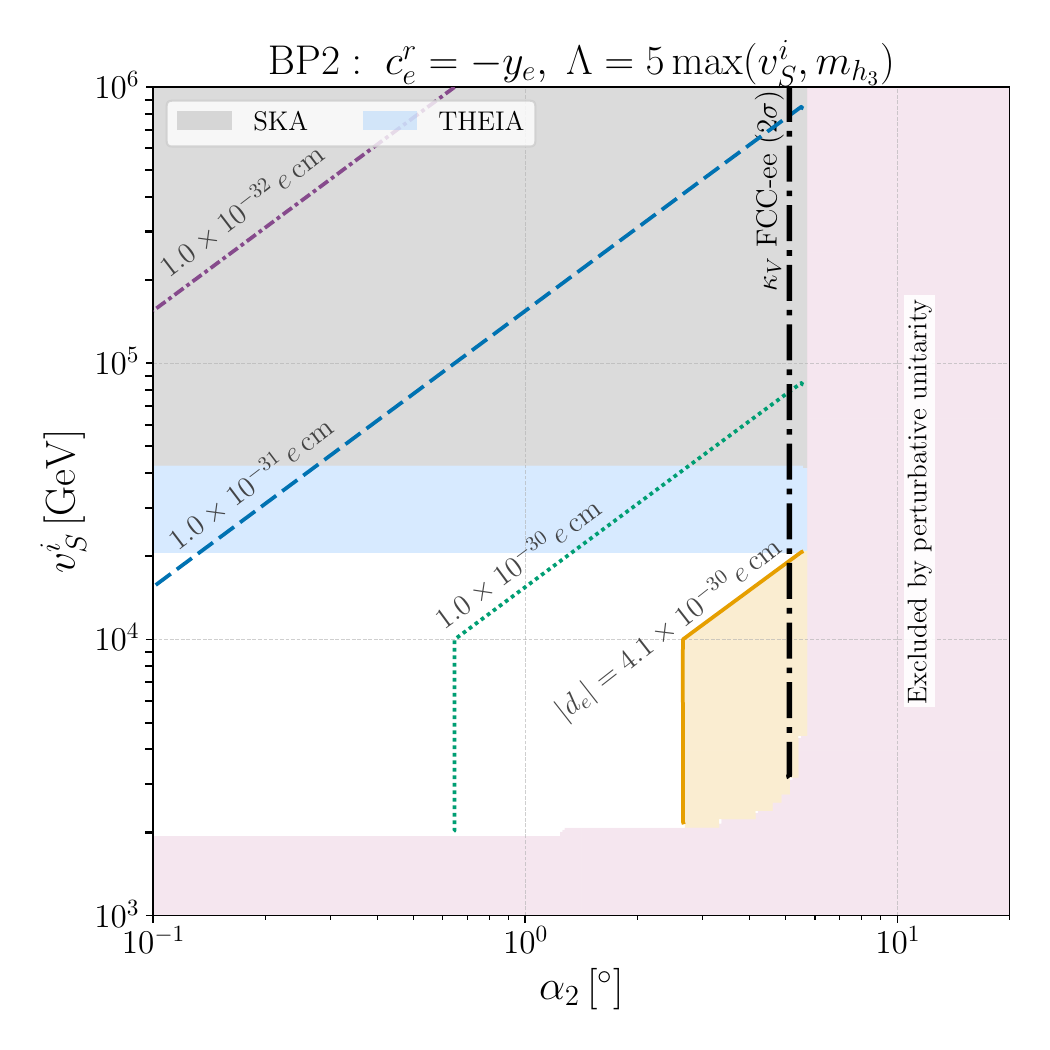} \\[1cm]
\includegraphics[width=7.5cm]{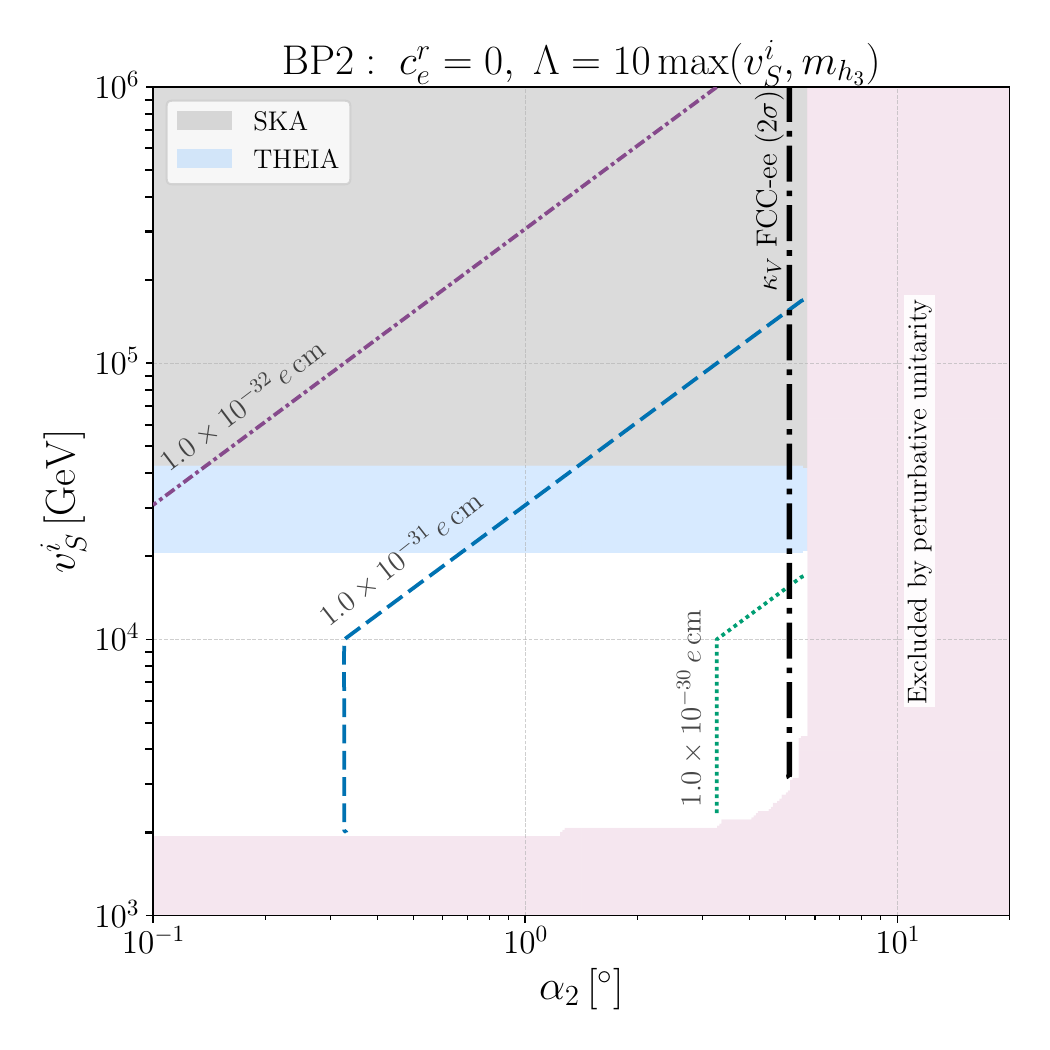} \hspace{0.5cm}
\includegraphics[width=7.5cm]{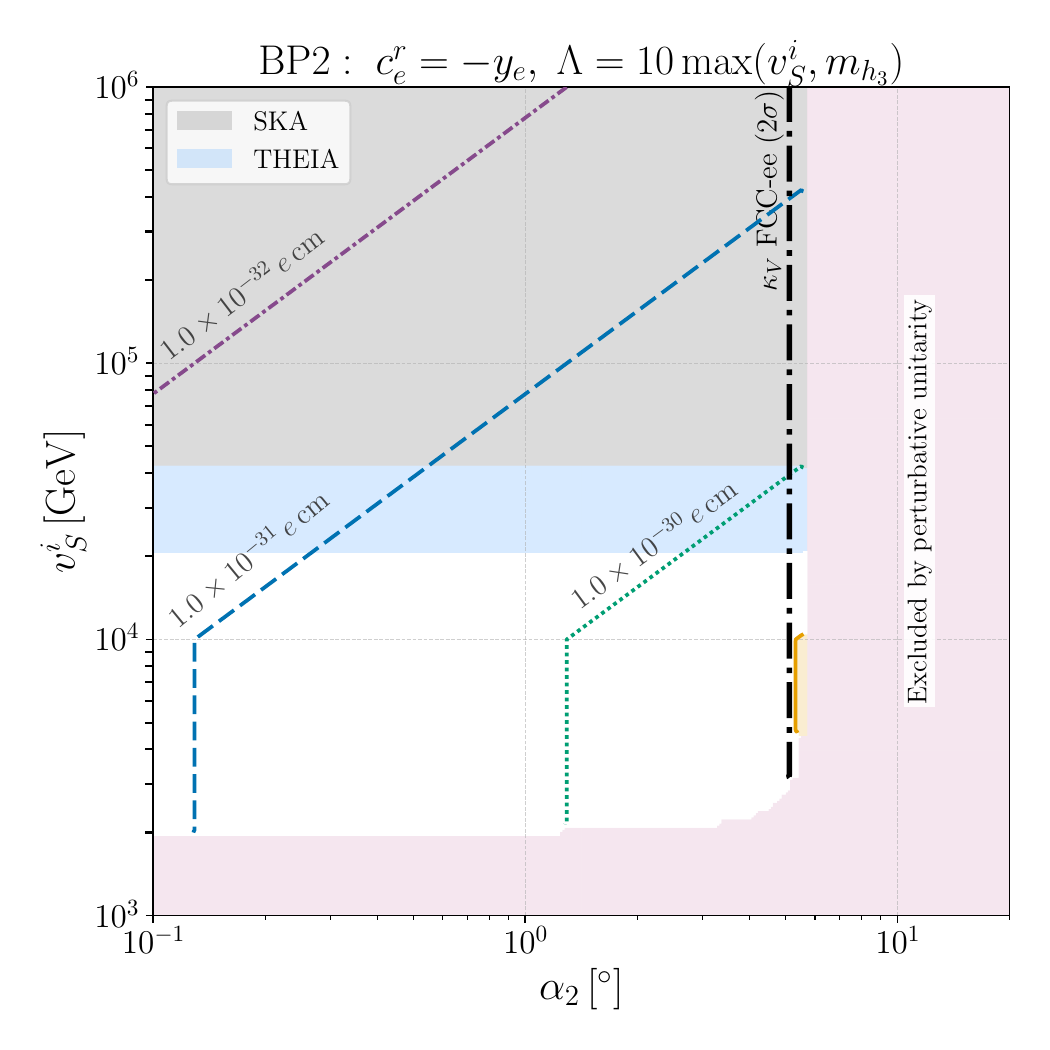}
\caption{
Contours of the electron EDM and the regions where the GW signal is detectable by SKA and THEIA are shown in the $(\alpha_2, v_S^i)$ plane for BP2 with $\Lambda = (5,10)\times\mathrm{max}(v_S^i,m_{h_3})$. 
The light magenta shaded region is excluded by the perturbative unitarity. 
The color scheme and line styles follow those in Fig.~\ref{fig:GW-de_alp2_vSi_bm1}.
}
\label{fig:GW-de_alp2_vSi_bm2}
\end{figure}

We now discuss the results for BP2 shown in Fig.~\ref{fig:GW-de_alp2_vSi_bm2}. 
In this case, the perturbative unitarity bound becomes relevant, corresponding to the light magenta shaded region with $\alpha_2\gtrsim 6^\circ$ and $v_S^i\lesssim 2~\mathrm{TeV}$. 
The color scheme and line styles follow those in Fig.~\ref{fig:GW-de_alp2_vSi_bm1}. 
Compared to BP1, the GW-detectable regions for SKA and THEIA are enlarged due to the larger value of $m_{h_3}$.

We comment on the parameter region with larger scalar masses and smaller $\alpha_2$. 
If $m_{h_2}$ and $m_{h_3}$ are larger than 10~TeV, the GW-detectable region expands due to the scaling $\hat{\sigma}_\mathrm{DW}\propto m_{h_3}^2$. 
However, the perturbative unitarity bound reduces the allowed range of $\alpha_2$ and shifts the lower bound on $v_S^i$ to larger values. 
The electron EDM scales as $d_e\propto 1/\mathrm{max}(v_S^i, m_{h_3})$, and therefore the region with $|d_e|<10^{-32}$ is enlarged. 

In contrast, in the region $\alpha_2<0.1^\circ$, the GW-detectable region remains essentially unchanged, while the region with $|d_e|<10^{-32}$ is broadened due to the scaling $d_e\propto \alpha_2$. 
Nevertheless, further improvements in the electron EDM sensitivity down to the $10^{-33}~\mathrm{e\,cm}$ level could probe such a region.

We next consider the fixed-cutoff scheme, choosing $\Lambda=5\times10^6~\mathrm{GeV}$ as a representative value. For both benchmarks, this choice keeps the cutoff at least five times larger than $\max(v_S^i,m_{h_3})$ throughout the scan up to $v_S^i=10^6~\mathrm{GeV}$. 
Compared with the prescription $\Lambda=5\max(v_S^i,m_{h_3})$, the fixed cutoff introduces a larger hierarchy between the cutoff and the relevant model scales whenever $\max(v_S^i,m_{h_3})<10^6~\mathrm{GeV}$. 
Consequently, the electron EDM induced by the dimension-five Yukawa interactions is further suppressed, whereas the GW-detectable regions remain unchanged at the accuracy of our calculation.

\begin{figure}[t]
\center
\includegraphics[width=7.5cm]{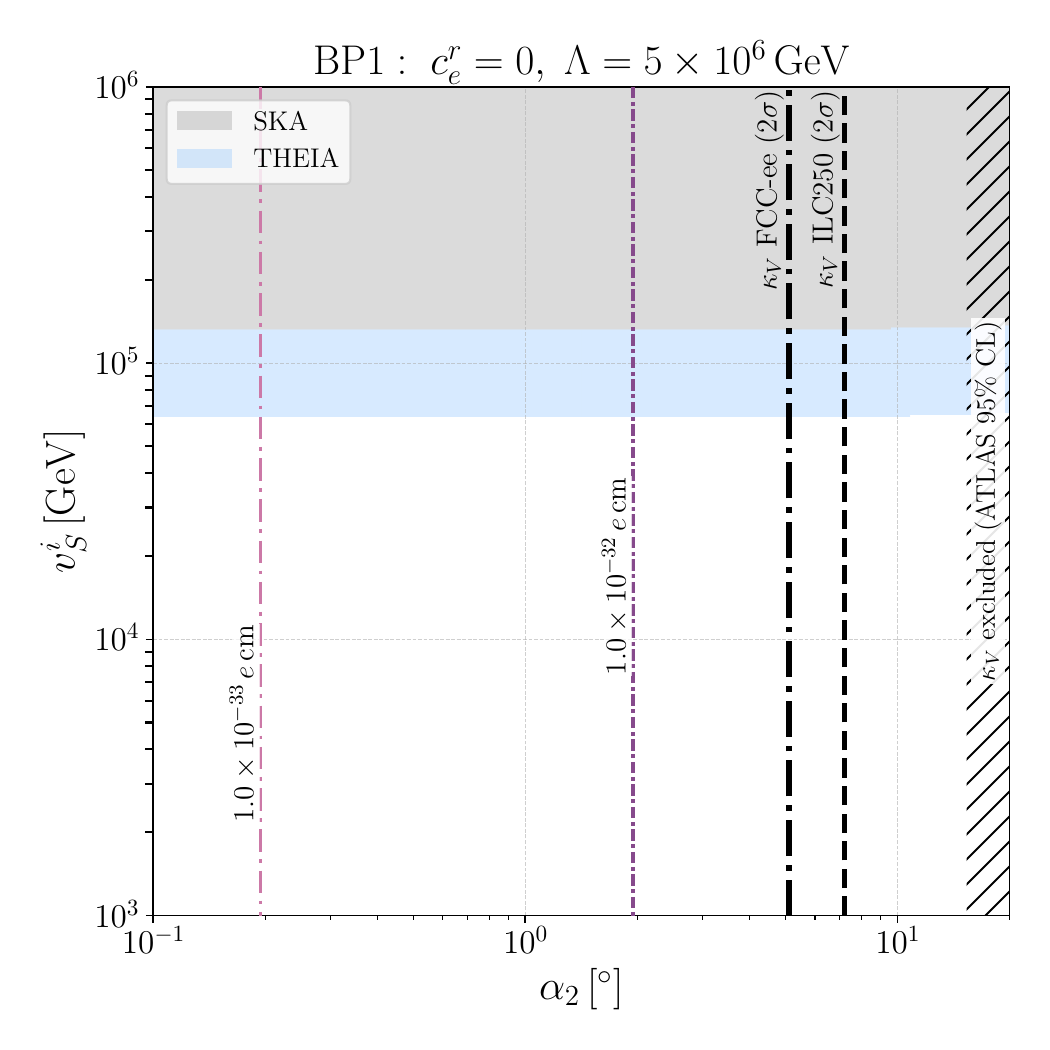}  \hspace{0.5cm}
\includegraphics[width=7.5cm]{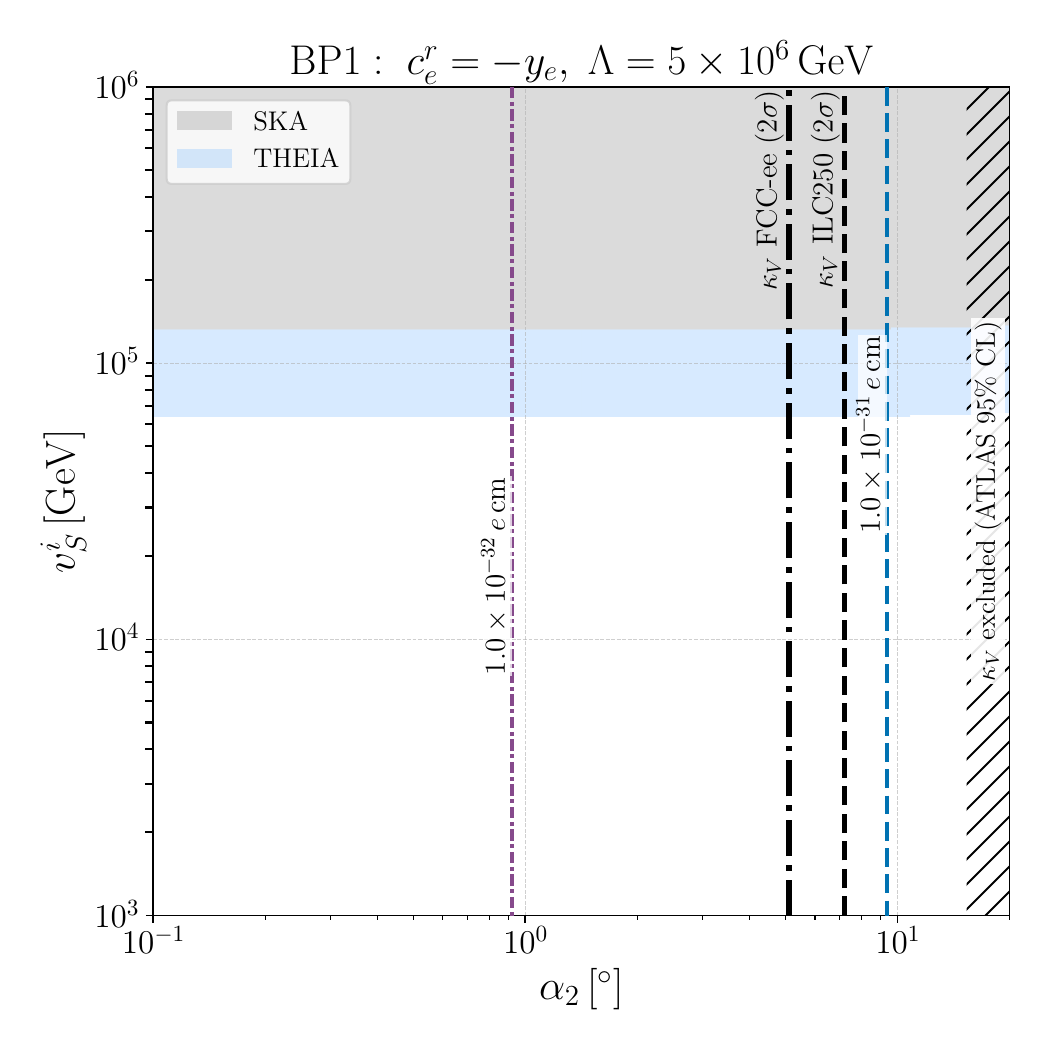} \\[1cm]
\caption{
Results for BP1 with $\Lambda=5\times10^6~\mathrm{GeV}$. The color scheme and line styles are the same as those in Fig.~\ref{fig:GW-de_alp2_vSi_bm1}, with an additional magenta dot-long-dashed line indicating $|d_e|=1.0\times10^{-33}~e\,\mathrm{cm}$.
}
\label{fig:GW-de_alp2_vSi_bm1p1}
\end{figure}

Fig.~\ref{fig:GW-de_alp2_vSi_bm1p1} shows the results for BP1 with $\Lambda=5\times10^6~\mathrm{GeV}$, taking $c_e^r=0$ in the left panel and $c_e^r=-y_e$ in the right panel. In addition to the contours shown previously, we display the contour $|d_e|=1.0\times10^{-33}~e\,\mathrm{cm}$ (dot-long-dashed, magenta). The remaining colors and line styles follow those in Fig.~\ref{fig:GW-de_alp2_vSi_bm1}. 
With $\Lambda$ held fixed, the approximate scaling $d_e\propto\alpha_2/\Lambda$ yields nearly vertical EDM contours in the $(\alpha_2,v_S^i)$ plane. 
The current electron EDM bound does not exclude any of the parameter space shown. 
A future sensitivity of $1.0\times10^{-31}~e\,\mathrm{cm}$ would probe part of the GW-detectable region for $c_e^r=-y_e$, owing to the constructive contributions discussed in Sec.~\ref{sec:edm}, but would not reach the corresponding region for $c_e^r=0$. 
Improving the sensitivity to the $10^{-32}~e\,\mathrm{cm}$ level would allow the GW-detectable region to be probed in both cases.

\begin{figure}[t]
\center
\includegraphics[width=7.5cm]{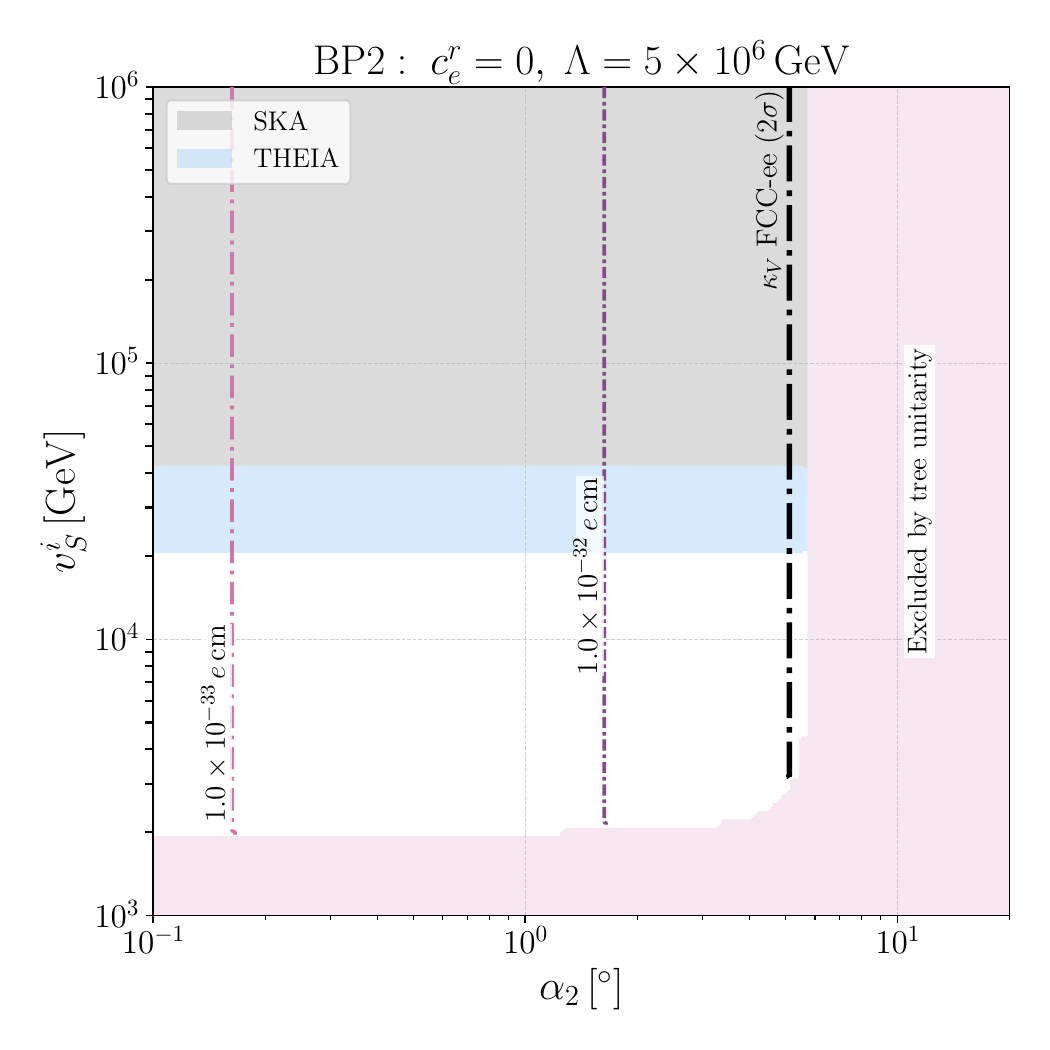}  \hspace{0.5cm}
\includegraphics[width=7.5cm]{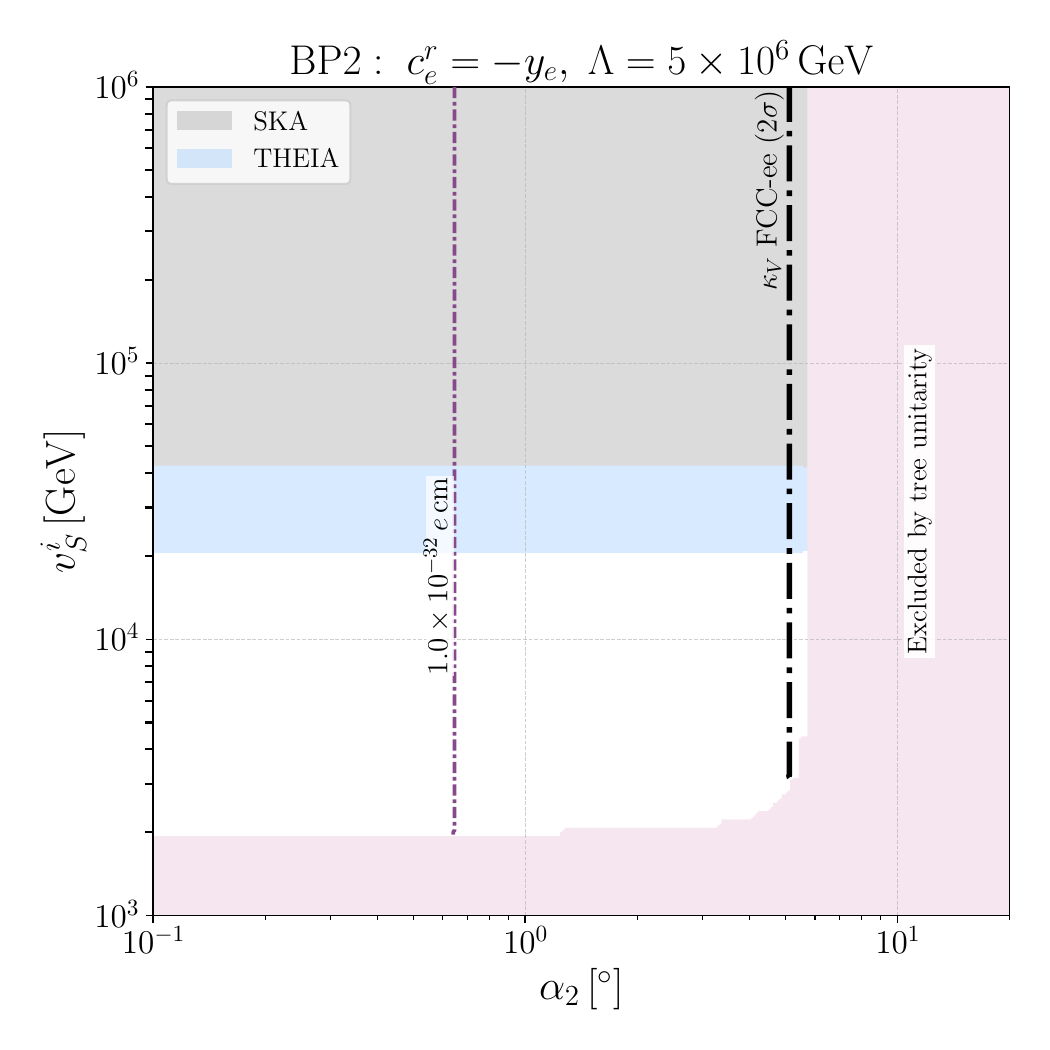} \\[1cm]
\caption{
Results for BP2 with $\Lambda=5\times10^6~\mathrm{GeV}$. The color scheme and line styles are the same as those in Figs.~\ref{fig:GW-de_alp2_vSi_bm2} and \ref{fig:GW-de_alp2_vSi_bm1p1}.
}
\label{fig:GW-de_alp2_vSi_bm2p1}
\end{figure}

Fig.~\ref{fig:GW-de_alp2_vSi_bm2p1} presents the corresponding results for BP2 with $\Lambda=5\times10^6~\mathrm{GeV}$ and $c_e^r=0$ (left) or $c_e^r=-y_e$ (right). 
The colors and line styles follow those in Figs.~\ref{fig:GW-de_alp2_vSi_bm2} and \ref{fig:GW-de_alp2_vSi_bm1p1}. 
For this benchmark, the heavier scalar spectrum imposes a stronger perturbative unitarity constraint on $\alpha_2$, limiting the electron EDM attainable in the allowed parameter space. 
A sensitivity of $1.0\times10^{-31}~e\,\mathrm{cm}$ would therefore be insufficient to probe the allowed GW-detectable regions in either case. 
Nevertheless, sensitivities at the $10^{-32}~e\,\mathrm{cm}$ level and below would probe portions of these regions, illustrating the complementarity between GW observations and electron EDM searches also in the fixed-cutoff scheme.

In summary, our demonstration here shows that the combined observation of GWs and EDM effects offers complementary probes of CPV in the singlet scalar sector and supports the interpretation that the GW signal may arise from CPDW collapses.

\section{Conclusion}\label{sec:conclusion}
We have investigated the interplay between stochastic GWs arising from the annihilation of CPDWs and the electron EDM in the cxSM with dimension-five Yukawa interactions. 
To clarify the CP structure of the extended model, we constructed basis-invariant CP-odd quantities that characterize both explicit CPV and SCPV. 
We then derived CPDW configurations associated with spontaneous CP breaking in the scalar sector and studied the interplay between GW signals from DW annihilation and the electron EDM. 
In particular, we examined whether the parameter region accessible to pulsar timing arrays, such as SKA, and space-based detectors, such as THEIA, can be probed by current or future EDM experiments.

Our analysis shows that the current electron EDM bound already excludes the parameter region in which the imaginary part of the singlet scalar VEV is below approximately $\mathcal{O}(10)$~TeV, near the maximal Higgs mixing allowed by current experiments, within our cutoff prescription.
In contrast, the parameter region relevant for GW detection by SKA and THEIA corresponds to larger values of the singlet VEV, typically above $\mathcal{O}(10)$~TeV. 
We find that a large fraction of this region can be probed by future EDM experiments with sensitivities at the $10^{-32}$--$10^{-31}~e\,\mathrm{cm}$ level, under mild assumptions on the dimension-five Yukawa couplings and the Higgs--gauge boson couplings.
This complementarity between cosmological and precision probes provides a novel strategy to explore the singlet scalar sector, linking its vacuum structure to observable CPV effects.

As discussed in Sec.~\ref{sec:gw}, the precise GW normalization and spectral shape from collapsing CPDWs still carry theoretical uncertainties associated with the annihilation dynamics.
Nevertheless, the qualitative GW--eEDM complementarity discussed in this work remains largely unchanged, since the main conclusion is based on the overlap between the parameter region relevant for observable GWs and that probed by future electron EDM experiments, rather than on the detailed GW spectrum itself.
Reducing the theoretical uncertainties in the GW prediction through improved simulations and analytical understanding of DW collapse dynamics will be important for more quantitative interpretations of future GW observations.

\begin{acknowledgments}
This research is funded by the Vietnam National Foundation for Science and Technology Development (NAFOSTED) under the grant number 103.01-2025.04.
\end{acknowledgments}
\appendix

\section{Basis-invariant CP phases}\label{basisCPV}
We construct the CP-odd basis invariants in the cxSM with the dimension-five Yukawa interactions.
The extended Yukawa sector is defined as
\begin{align}
-\mathcal{L}_Y &= \bar{q}_{L}\tilde{H}\left(Y^u+\frac{C^u}{\Lambda}S\right)u_{R}
+ \bar{q}_{L}H\left(Y^d+\frac{C^d}{\Lambda}S\right)d_{R} +\bar{\ell}_{L}H\left(Y^e+\frac{C^e}{\Lambda}S\right)e_{R}+{\rm H.c.}.
\end{align}
The basis transformations for $q_L$, $u_R$, $d_R$, $\ell_L$, and $e_R$ are, respectively, given by\footnote{$V_L^q$, $V_L^\ell$ are unitarity matrices for the doublet fields, $q_L = (u_L, d_L)^T$ and $\ell_L = (\nu_L, e_L)^T$.}
\begin{align}
q_L &= V_L^qq_L',\quad u_R= V_R^uu_R',\quad d_R = V_R^dd_R',\label{trf_quarks} \\
\ell_L &= V_L^\ell \ell_L',\quad e_R = V_R^ee_R'.
\label{trf_leptons}
\end{align}
The phase transformations of $H$ and $S$ are defined as
\begin{align}
H = e^{i\theta_H}H',\quad S = e^{i\theta_S}S'.
\label{trf_scalars}
\end{align}
Under the basis transformations for the fermions and scalars, the Yukawa and scalar couplings $Y^f$, $C^f$, $a_1$, and $b_1$ transform as
\begin{align}
Y^{u'} &= e^{-i\theta_{H}}V_L^{q\dagger}Y^uV_R^u, \quad Y^{d'} = e^{i\theta_{H}}V_L^{q\dagger}Y^dV_R^d, \quad Y^{e'} = e^{i\theta_{H}}V_L^{\ell \dagger}Y^eV_R^e, \label{newY}\\
C^{u'} &= e^{-i(\theta_{H}-\theta_S)}V_L^{q\dagger}C^uV_R^u,\quad 
C^{d'} = e^{i(\theta_{H}+\theta_S)}V_L^{q\dagger}C^dV_R^d,\quad
C^{e'} = e^{i(\theta_{H}+\theta_S)}V_L^{\ell \dagger}C^eV_R^e, \label{newC} \\
a_1' &= e^{i\theta_S}a_1,\quad b_1'  = e^{2i\theta_S}b_1,\quad v = e^{i\theta_H}v', \quad v_S = e^{i\theta_S}v_S'.\label{newSCoup}
\end{align}
The CP-odd basis invariant for the scalar sector is 
\begin{align}
I_1 = \mathrm{Im}(a_1^2b_1^*) \equiv |a_1^2||b_1|\sin\theta_1.
\end{align}
For each fermion sector $f=u,d,e$, $Y^f$ can be diagonalized by bi-unitary transformations,
\begin{align}
V_L^{f\dagger}Y^f_\mathrm{SM} V_R^f=Y^f_\mathrm{diag}=\mathrm{diag}(y_1^f, y_2^f, y_3^f),
\end{align}
where
\begin{align}
Y^f_\mathrm{SM}  = Y^f +\frac{C^fv_S}{\sqrt{2}\Lambda}.
\end{align}
Note that $Y_\mathrm{SM}^f$ has the same transformation law as that of $Y^f$.
Let us define the coupling $C^f$ in the mass eigenbasis as
\begin{align}
V_L^{f\dagger}C^f V_R^f \equiv c^f,
\end{align}
where $c^f$ are three-by-three complex matrices.

One of the CP-odd basis invariant including $Y_\mathrm{SM}^f$ is given by
\begin{align}
I^f &= \mathrm{Im}\mathrm{Tr}(a_1Y_\mathrm{SM}^fC^{f\dagger}) = \mathrm{Im}\mathrm{Tr}(a_1Y_\mathrm{diag}^f c^{f\dagger}) = \mathrm{Im}\left[a_1\sum_{i=1}^3(y_i^fc_{ii}^{f*})\right],
\end{align}
where $\mathrm{Tr}$ is defined on the generation space. 
Note that the off-diagonal elements do not appear in $I^f$.
Hence, $I^f$ is the measure of CPV for the diagonal elements of $c^f$.

We now turn to the CP-odd invariants involving the singlet VEV $v_S$. Under the field rephasing in Eq.~\eqref{newSCoup}, the phases of $a_1$, $b_1$, and $v_S$ transform as
\begin{align}
\theta'_{a_1} &= \theta_{a_1}+\theta_S, \quad
\theta'_{b_1} = \theta_{b_1}+2\theta_S, \quad
\theta'_{v_S} = \theta_{v_S}-\theta_S.
\end{align}
The combinations $\phi_1 \equiv \theta_{a_1}+\theta_{v_S}$ and $\phi_2 \equiv \theta_{b_1}+2\theta_{v_S}$ are therefore invariant under this rephasing. We can accordingly define the CP-odd invariants
\begin{align}
J_a &= \mathrm{Im}(a_1v_S)= |a_1||v_S|\sin\phi_1, \\
J_b &= \mathrm{Im}(b_1v_S^{2})= |b_1||v_S|^2\sin\phi_2= |b_1||v_S|^2\sin(2\phi_1-\theta_1).
\end{align}
The invariants $J_a$ and $J_b$ characterize CPV involving the vacuum phase, in contrast to $I_1$, which depends only on the parameters of the scalar potential. 
In the absence of explicit CPV, the scalar potential can be made real by a field rephasing. 
A nonzero value of $J_a$ then signals SCPV. Furthermore, the tadpole condition relates $J_a$ and $J_b$ through $\sqrt{2}J_a+J_b/2=0$, as shown in Eq.~\eqref{scpv_J}, so they are not independent at a stationary point. 
Consequently, for nonzero $a_1$ and $b_1$, it is sufficient to use $J_a$ to characterize SCPV in the scalar sector.

Moreover, as a dependent parameter, we define another CP-odd invariants, including $v_S$ as
\begin{align}
J_f&=\mathrm{Im}\mathrm{Tr}\Big[v_S (Y_\mathrm{SM}^{f})^{-1} C^f\Big] 
= \mathrm{Im}\mathrm{Tr}\Big[v_S(Y_\mathrm{diag}^{f})^{-1}c^{f}\Big] = \mathrm{Im}\left[v_S\sum_{i=1}^3\frac{c_{ii}^{f}}{y_i^{f}}\right].
\end{align}
In our simplified setup, the invariants are reduced to
\begin{align}
J_f  =  \mathrm{Im}\left[v_S\frac{c_{f}}{y_f} \right],
\end{align}
where $f=t,e$.
It should be noted that
\begin{align}
\mathrm{Arg}\left(v_S\frac{c_{t}}{y_t}\right) & = \theta_{v_S}+\theta_{c_t}-\theta_{y_t}  = \phi_1-\theta_2,\\
\mathrm{Arg}\left(v_S\frac{c_{e}}{y_e}\right) &= \theta_{v_S}+\theta_{c_e}-\theta_{y_e}= \phi_1-\theta_3.
\end{align}
Thus, the CPV phases of $J_f$ are not independent. 
Although $J_f$ are CP-odd basis invariants, the electron EDM is not necessarily expressed in terms of them. 
This is because the transformation between $(h, s, \chi)$ and $(h_1, h_2, h_3)$ is not taken into account in defining $J_f$. 
This is the reason why the electron EDM has the form of Eq.~(\ref{de_an}).
In the only limited case, the electron EDM can be expressed in terms of $J_t$ and $J_e$, as shown in Eq.~(\ref{de_an_lim}).

\section{Electron EDM}
The expressions for $d_e^{h\gamma}$ and $d_e^{hZ}$ defined in Eq.~(\ref{de_BZ}) are, respectively, given
 by~\cite{Ellis:2008zy,Cheung:2009fc,West:1993tk,Chang:2005ac,Ellis:2010xm,Cheung:2014oaa,Inoue:2014nva,Bowser-Chao:1997kjp,Abe:2013qla}
\begin{align}
\frac{(d_{e}^{h\gamma})_t}{e}
&= \frac{\alpha_{\rm em}}{6\pi^3m_t}\sum_{i=1}^3
\Big[
	g_{h_i\bar{e}e}^Pg_{h_i\bar{t}t}^Sf(\tau_{th_i})
	+g_{h_i\bar{e}e}^Sg_{h_i\bar{t}t}^Pg(\tau_{th_i})
\Big], \label{de_hgam_t}\\
\frac{(d_{e}^{h\gamma})_W}{e}
& = -\frac{\sqrt{2}G_f\alpha_\mathrm{em}v}{32\pi^3}\sum_{i=1}^3g_{h_i\bar{e}e}^Pg_{h_iVV}\mathcal{J}_W^\gamma(m_{h_i}), \label{de_hgam_W}\\
\frac{(d_{e}^{hZ})_t}{e} 
& = -\sum_{i=1}^3\frac{\alpha_{\text{em}}v_{Z\bar{e}e}v_{Z\bar{t}t}}{4\pi^3m_fs_W^2c_W^2}
\Big[g_{h_i\bar{e}e}^Sg_{h_i\bar{t}t}^P\tilde{F}_A(\tau_{th_i},\tau_{tZ})+g_{h_i\bar{e}e}^Pg_{h_i\bar{t}t}^S\tilde{F}_H(\tau_{th_i},\tau_{tZ}) \Big], \\
\frac{(d_{e}^{hZ})_W}{e}
&=\sum_{i=1}^3 \frac{\sqrt{2}\alpha_{\text{em}}G_Fvv_{Z\bar{e}e}}{32\pi^3s_W^2} g_{h_i \bar{e}e}^Pg_{h_i VV}^{}\mathcal{J}^Z_W(m_{h_i}), 
\end{align}
where $\alpha_\mathrm{em}=e^2/4\pi$ with $e$ being the positron charge, $\tau_{AB}=m_A^2/m_B^2$, $s_W=\sin\theta_W$, and $c_W=\cos\theta_W$.
The $Z$-fermion coupling is defined by $v_{Z\bar{f}f}=T_f^3/2-Q_fs_W^2$, where $T_f^3=\pm 1/2$ and $Q_f$ denote the third component of weak isospin and the electric charge of the fermion $f$, respectively.
The loop functions are summarized in Ref.~\cite{Nhi:2025iob}.

For $d_e^\mathrm{kite}$, we use the expressions given in Ref.~\cite{Nhi:2025iob}, whose notation matches ours. 
The original derivation was presented in Ref.~\cite{Altmannshofer:2020shb}.

%
\bibliography{refs}
%

\end{document}